\documentclass[a4paper,12pt]{article}

\usepackage{amsmath,amssymb,amsthm,mathrsfs}
\usepackage{latexsym}
\usepackage[dvips]{graphicx}
\newcommand{\del}{\partial}
\newcommand{\Tr}{\mathrm{Tr}}
\newcommand{\Xdot}{\dot{X}}
\newcommand{\Rt}{\tilde{R}}
\newcommand{\Ocal}{\mathcal{O}}
\newcommand{\gym}{g_{\mathrm{YM}}}

\newcommand{\phit}{\tilde{\phi}}

\newcommand{\phihat}{\hat{\phi}}
\newcommand{\Ahat}{\hat{A}}

\newcommand{\Fhat}{\hat{F}}
\newcommand{\omegat}{\tilde{\omega}}
\newcommand{\ubar}{\bar{u}}
\newcommand{\ahat}{\hat{a}}
\newcommand{\bhat}{\hat{b}}
\newcommand{\abf}{\mathbf{a}}
\newcommand{\bbf}{\mathbf{b}}
\newcommand{\Mbf}{\mathbf{M}}
\newcommand{\Lambdabf}{\mathbf{\Lambda}}
\newcommand{\Ubf}{\mathbf{U}}
\newcommand{\gbar}{\bar{g}}
\newcommand{\Abar}{\bar{A}}
\newcommand{\phibar}{\bar{\phi}}
\newcommand{\Ctilde}{\tilde{C}}
\newcommand{\Fscr}{\mathscr{F}}
\newcommand{\Gscr}{\mathscr{G}}
\newcommand{\Ftilde}{\tilde{F}}
\newcommand{\Gtilde}{\tilde{G}}
\newcommand{\Htilde}{\tilde{H}}
\newcommand{\Btilde}{\tilde{B}}
\newcommand{\bbar}{\bar{b}}
\newcommand{\abar}{\bar{a}}
\newcommand{\frakw}{\mathfrak{w}}
\newcommand{\Fcal}{\mathcal{F}}
\newcommand{\Tcal}{\mathcal{T}}
\newcommand{\Jcal}{\mathcal{J}}
\newcommand{\Hscr}{\mathscr{H}}
\newcommand{\ebar}{\bar{\epsilon}}
\newcommand{\pbar}{\bar{p}}
\newcommand{\nbar}{\bar{n}}
\newcommand{\nablabar}{\bar{\nabla}}

\newcommand{\Dcal}{\mathcal{D}}
\newcommand{\xhat}{\hat{x}}
\newcommand{\khat}{\hat{k}}
\newcommand{\rhat}{\hat{r}}
\newcommand{\That}{\hat{T}}

\begin{document}
\renewcommand{\thefootnote}{\fnsymbol{footnote}}
\begin{titlepage}
\hfill
{\hfill \begin{flushright} 
KEK-TH-1630
\end{flushright}
}

\vspace*{10mm}

\begin{center}
{\Large \textbf{
Linear responses of D0-branes   via  gauge/gravity correspondence }}
\vspace*{12mm}

\normalsize{Yoshinori Matsuo$^{1}$\footnote{E-mail: ymatsuo@post.kek.jp}, Yuya Sasai$^2$\footnote{E-mail: sasai@law.meijigakuin.ac.jp}, Yasuhiro Sekino$^{1}$\footnote{E-mail: sekino@post.kek.jp}}

\vspace*{12mm}
\textit{
$^1$ KEK Theory Center, High Energy Accelerator Research Organization (KEK), Tsukuba 305-0801, Japan \\
$^2$ Institute of Physics, Meiji Gakuin University, Yokohama 244-8539, Japan  
}

\end{center}

\vspace*{12mm}

\begin{abstract}
We study  linear responses of   D0-branes in the low frequency region by using gauge/gravity correspondence. The dynamics of the D0-branes is described by Matrix theory with finite temperature, which is dual to a near extremal D0-brane black hole solution. We analyze the tensor mode and  vector modes of a  stress tensor and  a  Ramond-Ramond 1-form current of Matrix theory.  Then, we show that if a cutoff surface  is close to a horizon of the D0-brane black hole, the  linear responses take  forms   similar to the hydrodynamic stress tensor and current on $S^8$. By taking a Rindler limit, those linear responses come to obey the hydrodynamics exactly, which is consistent with  previous works on a Rindler fluid.  We also show that if the cutoff surface is far from the horizon, the linear responses do not take the forms of the hydrodynamic stress tensor and  current on $S^8$. Especially, we find that the vector modes no longer possess a diffusion pole  in the low frequency region, which indicates that the linear responses of the D0-branes cannot be explained by hydrodynamics.
\end{abstract}

\end{titlepage}

\newpage
\renewcommand{\thefootnote}{\arabic{footnote}}
\setcounter{footnote}{0}

\section{Introduction}
If we apply a time-dependent external field to a black hole, what occurs
in the black hole?  According to the membrane paradigm
\cite{Damour:1978cg,Thorne:1986iy,Parikh:1997ma}, the response of the
black hole can be represented by the degrees of freedom
on the stretched horizon. It is expected that these degrees of freedom
carry information on the interior of the black hole. This is 
suggested by two guiding principles of quantum gravity: the
holographic principle (which states that the entropy in a spatial
region is bounded by the area) and the black hole
complementarity (which states that there is a consistent theory
in the frame of an observer outside the horizon).

It has long been known that matter on the stretched horizon
obeys hydrodynamic laws in the 
long wavelength limit~\cite{Damour:1978cg}. On the other hand,
it was pointed out that a localized perturbation spreads 
over the entire horizon in a time logarithmic in the 
Bekenstein-Hawking entropy. This time scale, which is
different from the ones in local quantum field theories, 
plays a crucial role in forbidding a possible violation of 
 the no-cloning theorem of quantum state~\cite{Hayden:2007cs,
Sekino:2008he,Susskind:2011ap}. These properties have
been derived from classical general relativity. 
To understand what the degrees of freedom on the stretched
horizon are, and how they thermalize, we will need a fundamental theory, 
such as string theory.
 

In string theory, the Bekenstein-Hawking entropy and the Hawking emission
rate of some specific black holes have been correctly reproduced by 
D-brane systems~\cite{Strominger:1996sh,Callan:1996dv}. It is likely 
that D-branes provide microscopic descriptions 
for more general black holes,
but there are few quantitative results.


Recently, it has  been  found that  transport coefficients in the
membrane paradigm agree with those of a highly excited fundamental
string at the correspondence point, up to numerical coefficients
\cite{Sasai:2010pz,Sasai:2011ys}. This can be regarded as a
support for string theory as a microscopic description of the
stretched horizon, in spite of a few limitations.
First, the black hole is realized when the string coupling is larger
than the value at the correspondence point \cite{Horowitz:1996nw}. In
this situation, one can no longer neglect the excitations of D-branes
because the masses of the D-branes are proportional to the inverse of
the string coupling  \cite{Mathur:1997wb}.\footnote{In
\cite{Sasai:2011hw}, transport coefficients of the D1-D5-P system induced by
a few moduli fields  have been discussed. However, since the low energy
effective theory of the D1-D5-P system does not couple to the bulk
metric and gauge field \cite{Callan:1996dv,David:2002wn}, we could not
discuss the linear responses of the  stress tensor and current.} In
addition, the authors of \cite{Sasai:2010pz,Sasai:2011ys} have considered only 
homogeneous perturbations to obtain the transport coefficients of the fundamental string. 
To study  dynamical processes such as diffusion, one has to apply inhomogeneous perturbations to the system.


According to AdS/CFT correspondence or Matrix theory, 
string theory can be defined nonperturbatively by 
supersymmetric Yang-Mills theories. These gauge theories
should allow us to study the dynamics on the stretched 
horizon from the first principles, even though it is 
difficult to solve these theories. 
In fact, there have been extensive studies of  
hydrodynamic properties of strongly coupled gauge 
theories from gravity calculations using
AdS/CFT correspondence (or 
gauge/gravity correspondence, more generally), 
following the work of Policastro \textit{et al.}
\cite{Policastro:2002se}. 
In these studies, transport coefficients of gauge theories
have been obtained by studying fluctuations around black brane 
backgrounds which have momentum along the brane, in the limit
of small momentum. The computations of gauge/gravity 
correspondence are closely related to those in the old membrane 
paradigm~\cite{Bredberg:2010ky}. 

In spite of this development, it is not clear how 
the transport phenomena in the directions which surround 
a black hole (or a black brane; the $S^5$ direction
in the case of D3-branes) are represented in gauge theories. 
This question should be important in the understanding of black holes in
the real world, since there are no directions along the brane in this
case. Also, answers to this question may shed light on how the space emerges 
from lower dimensional theories. 

Let us consider the case of D0-branes
for definiteness. There are no spatial directions along 
the brane, and the D0-brane black hole is surrounded by $S^8$. 
The low-energy description of D0-branes is given by maximally 
supersymmetric 
(0+1) dimensional Yang-Mills theory with $U(N)$ gauge symmetry.
This theory is called Matrix theory, and has been proposed to be a 
description of M theory in a particular
large $N$ limit~\cite{Banks:1996vh}.

Black holes should correspond to dynamically realized
spherically symmetric configurations of matrix-valued scalar fields.
Fluctuations on the stretched
horizon should correspond to fluctuations around such a 
configuration. It is not clear how these fluctuations propagate,
and it is not even clear if they  can be effectively described by
a local field theory on $S^8$.\footnote{In \cite{Asplund:2011qj,Asplund:2012tg}, fluctuations of  Matrix theory have been analyzed by  a numerical simulation of the classical dynamics to study the  thermalization in the high temperature regime.}

An important clue is that one knows how the matrices couple
to  background fields. Kabat and Taylor~\cite{Kabat:1997sa} and 
Taylor and Van Raamsdonk~\cite{Taylor:1998tv, Taylor:1999gq, Taylor:1999pr} 
studied one-loop effective potential in Matrix theory and
found that certain single-trace operators couple to
supergravity backgrounds. These operators have definite SO(9) R-charges, 
meaning that they are in the momentum representation
on $S^8$. One should be able to find linear responses of Matrix theory 
to external perturbations by computing correlation functions
of these operators. 

In this paper, we will study transport phenomena along $S^8$ 
in Matrix theory by using gauge/gravity correspondence. 
Our aim is to clarify what kind of behavior one should expect
from the dynamics of matrices. In particular, we wish to understand
to what extent the theory behaves as in field theory on $S^8$. 

Gauge/gravity correspondence for D0-branes was proposed
in \cite{Itzhaki:1998dd, Jevicki:1998yr}. Correlation functions
at zero temperature have been  
found~\cite{Sekino:1999av, Sekino:2000mg, Asano:2003xp, 
Asano:2004vj, Kanitscheider:2008kd} 
by applying the 
Gubser-Klebanov-Polyakov-Witten(GKPW)~\cite{Gubser:1998bc,Witten:1998qj} 
prescription to the near-horizon D0-brane background. 
It was found that the zero-temperature correlators for
operators which couple to supergravity modes obey power law, 
even though the theory is not 
conformally invariant. These results have been
confirmed by Monte Carlo simulations of 
Matrix theory~\cite{Hanada:2009ne, Hanada:2011fq}.

In this paper, we follow the standard procedure for
studying the hydrodynamic limit in gauge/gravity
correspondence. We will use the real-time prescription 
proposed by \cite{ Policastro:2002se,Son:2002sd}. 
We evaluate the on-shell action on the near-extremal
D0-brane background and obtain correlation 
functions following the GKPW prescription. We make
a series expansion in the frequency  
and study the low frequency limit.
We will study the tensor and vector modes and 
find shear viscosity and diffusion poles for the stress tensor and 
Ramond-Ramond (R-R) 1-form current. The scalar modes are deferred 
to future work. 

One should note that the types of operators that we consider
are different from the ones familiar in the holographic 
study of hydrodynamics. The stress-energy tensor on 
$S^8$ is represented by scalar operators from the 
perspective of gauge theory on (0+1) dimension. Modes with
different momentum on $S^8$ are represented by different
operators. Unlike stress-energy tensors in conformal field 
theories, these operators are not marginal operator, and 
will have nontrivial wave function renormalization. 

We follow the interpretation in 
\cite{Bredberg:2010ky,Susskind:1998dq, Heemskerk:2010hk} 
and assume that the position $r=r_c$ of the regulated boundary 
(or the ``cutoff surface''), on which the on-shell action 
is evaluated, sets the scale of renormalization.  
We assume that the normalization of the operators is 
fixed at that scale. Since the gauge theory has only time,
the renormalization scale refers to the scale of 
time separation. 

We will consider the two cases: when the cutoff surface is 
near the horizon and when it is near infinity. 
In the former case, we obtain the results which can be
interpreted as conventional hydrodynamics. This is the
limit where the operators are defined at an infrared
scale. However, in the latter case, we observe that the 
theory behaves differently from the usual fluid. This is the
limit where the operators are defined at some ultraviolet
scale, so that the operators could be sensitive to the
short-time behavior of the theory. 


This paper is organized as follows. 
In section \ref{sec:hydro}, we review hydrodynamic equations for 
a charged fluid on $S^8$. We consider the tensor and vector
modes, and we  find the expressions for the stress tensor and 
current in the presence of external perturbations.
In section 3,  after briefly reviewing Matrix theory and 
gauge/gravity correspondence at zero temperature, we describe
gauge/gravity correspondence at finite temperature on which our
analysis is based. 
In section \ref{sec:linearresponse}, we calculate the linear responses
of the stress tensor and R-R 1-form current of Matrix theory by using
the gauge/gravity correspondence.\footnote{In this paper, we assume that
the stress tensor of Matrix theory is coupled to the  mode of the metric
in the bulk and the R-R 1-form current of Matrix theory is coupled to
the  mode of the R-R 1-form field in the bulk. This is different from
the correspondence between the operators in Matrix theory and the modes
of the supergravity fields proposed in \cite{Sekino:1999av}. However, we
believe that our assumption is more natural to obtain the correct linear
responses of Matrix theory under the perturbations of the bulk metric
and  R-R 1-form field.}  
In section \ref{sec:tensor} and \ref{sec:vector}, we calculate the 
on-shell action for the tensor and vector modes, respectively,  
at arbitrary $r_c$. In section \ref{sec:uc1}, we study transport
coefficients when the cutoff surface is near the horizon. 
From the tensor mode, we find that the linear response of the stress 
tensor takes the form of the hydrodynamic stress tensor on $S^8$, 
and that the shear viscosity to entropy density 
ratio is equal to $1/4\pi$. 
From the vector modes, we find that the linear responses of 
the stress tensor and R-R 1-form current take forms similar to 
the hydrodynamic stress tensor and current on $S^8$. By taking 
a Rindler limit, the linear responses  become the hydrodynamic 
stress tensor and current on $\mathbf{R}^8$, which is consistent 
with  previous work on a Rindler fluid~\cite{Bredberg:2010ky,
Matsuo:2012pi}. 
In section \ref{sec:uc0}, we consider the case in which 
the cutoff surface is far from the horizon. We find that both 
the tensor mode and  vector modes do not follow the hydrodynamics. 
Especially, there is no diffusion pole in the vector modes in the 
low frequency region, which indicates that the linear responses 
of  the  D0-branes cannot be explained by hydrodynamics. 
The final section is devoted to the summary and comments.
In  Appendix \ref{sec:harmonics}, we briefly summarize the 
definitions and properties of the spherical harmonics on $S^8$.  
In  Appendix \ref{sec:on-shell}, we derive the on-shell action 
of the tensor mode and vector modes.

\section{Hydrodynamics on $S^8$} \label{sec:hydro}
In this section, we review a charged fluid on 9-dimensional spacetime whose spatial part is $S^8$. 
We introduce external perturbations of the metric $g_{\mu\nu}$ and gauge field $A_\mu$ 
and consider the linear response \cite{Kovtun:2012rj}. 
The background metric and gauge field are given by 
\begin{align}
\gbar_{\mu\nu}&=
\begin{pmatrix}
-1 & 0 \\
0 & \gbar_{ij}
\end{pmatrix}, \\
\Abar_{\mu}&=(\bar{\mu}, 0),
\end{align}
where  the indices $\mu, \nu$ run from $0$ to $8$, and $\gbar_{ij}$ is the metric on $S^8$ with radius $R$. 
We introduced the chemical potential $\mu$ as the constant mode of $A_0$, and $\bar\mu$ 
is its background part. 

The hydrodynamic equations of the charged fluid are
\begin{align}
0&=\nabla_{\mu}T^{\mu\nu}-F^{\nu\mu}J_{\mu}, \label{eq:hydroeqstr} \\
0&=\nabla_{\mu}J^{\mu},
\end{align}
where $F_{\mu\nu}=\del_{\mu}A_{\nu}-\del_{\nu}A_{\mu}$ is the field strength. The constitutive relations of the stress tensor and current are
\begin{align}
T^{\mu\nu}&=\epsilon u^{\mu}u^{\nu}+p\Delta^{\mu\nu} -\eta\Delta^{\mu\alpha}\Delta^{\nu\beta}(\nabla_{\alpha}u_{\beta}+\nabla_{\beta}u_{\alpha}-\frac{1}{4}g_{\alpha \beta}\nabla_{\gamma}u^{\gamma})-\zeta \Delta^{\mu\nu}\nabla_{\gamma}u^{\gamma}, \\
J^{\mu}&=nu^{\mu}+\sigma \Delta^{\mu\lambda}(E_{\lambda}-T\Delta_{\lambda\rho}\nabla^{\rho}(\mu/T)),
\end{align}
where $\epsilon$ is the energy density, $p$ is the pressure, $n$ is the charge density, 
$T$ is the temperature, $\mu$ is the chemical potential, and
\begin{align}
\Delta^{\mu\nu}&=g^{\mu\nu}+u^{\mu}u^{\nu}, \\
E_{\mu}&=F_{\mu\nu}u^{\nu}, 
\end{align}
are the projection to spatial direction and the external electric flux, respectively. 
The coefficients of second order parts, $\eta, \zeta$, and  $\sigma$, are the shear viscosity, bulk viscosity, and conductivity, respectively. The normalization condition of the velocity field $u^\mu$ is given by
$g_{\mu\nu}u^{\mu}u^{\nu}=-1$.

Now, we introduce perturbations for the metric and gauge field, and then
consider the response of the fluid at the linear order of perturbations. 
By expanding them in terms of spherical harmonics on $S^8$, 
they can be classified into the tensor, vector, and scalar modes which are associated to 
the tensor, vector, and scalar harmonics, respectively. 
We consider the tensor mode and vector modes and introduce no perturbation for the scalar mode. 
Then, the scalar quantities such as $\epsilon$,  $p$, and $n$ have no response and 
remain  constant. Since in this case, the velocity field  satisfies the incompressible condition $\nabla_{\mu}u^{\mu}=0$, the constitutive relations are simplified as
\begin{align}
T^{\mu\nu}&=\ebar u^{\mu}u^{\nu}+\pbar \Delta^{\mu\nu} -\eta\Delta^{\mu\alpha}\Delta^{\nu\beta}(\nabla_{\alpha}u_{\beta}+\nabla_{\beta}u_{\alpha}), \\
J^{\mu}&=\nbar u^{\mu}+\sigma \Delta^{\mu\nu}E_{\nu},
\end{align}
where $\ebar, \pbar$, and $\nbar$ denote the energy density, pressure, and charge density in equilibrium, respectively.

We apply  the following  external perturbations to the fluid in equilibrium:
\begin{align}
g_{\mu\nu}&=\gbar_{\mu\nu}+h_{\mu\nu}, \\
A_{\mu}&=\bar{A}_{\mu}+\delta A_{\mu},
\end{align}
where 
\begin{align}
h_{ij}(t,x^i)&=\sum_I b^I(t)Y^I_{ij}(x^i), \\
h_{0i}(t,x^i)&=\sum_Ib^I_0(t)Y^I_i(x^i), \\
\delta A_i(t,x^i)&=\sum_I a^I(t) Y^I_i(x^i).
\end{align}
Here, $Y^I_i$ and $Y^I_{ij}$ are the  vector harmonics and tensor harmonics on $S^8$, respectively. The tensor mode is $b^I$ and the vector modes are $b_0^I$ and $a^I$. Hereafter, we  often suppress the angular momentum index $I$ and the sum in the spherical  harmonic expansions. The definitions and properties of the spherical harmonics are summarized in  Appendix \ref{sec:harmonics}. 

Under the perturbations, the velocity field changes as
\begin{align}
u^{\mu}&=\ubar^{\mu}+\delta u^{\mu}, 
\end{align}
where $\ubar^\mu$ is the velocity field in the equilibrium and 
the linear responses can be expanded by the spherical harmonics: 
\begin{align}
\ubar^{\mu}&=(1,0,\cdots, 0), \\
\delta u^0&=0, ~~~~\delta u^i=u(t)Y^i. 
\end{align}
Since $u^0$ behaves as a scalar on $S^8$, it does not change in this case. 
Note that $\delta u_i=\delta u^{\mu}\gbar_{\mu i}+\ubar^{\mu} \delta g_{\mu i}=(u-b^0)Y_i$.
Thus, the changes of the stress tensor and  current are
\begin{align}
\delta T^{0i}&=((\ebar +\pbar) u-\pbar b^0)Y^i, \label{eq:dT0i} \\
\delta T^{ij}&=-(\pbar b+\eta \del_0 b)Y^{ij}-\eta u (\nablabar^i Y^j +\nablabar^j Y^i), \\
\delta J^i&=(\nbar u-\sigma \del_0 a)Y^i. \label{e:dJi}
\end{align}
Inserting (\ref{eq:dT0i})-(\ref{e:dJi}) into the hydrodynamic equation (\ref{eq:hydroeqstr}), we find
\begin{align}
u(\omega)&=\frac{i\omega  b^0(\omega)-i\omega \frac{\nbar}{\ebar +\pbar} a(\omega)}{i\omega -D \frac{(l+8)(l-1)}{R^2}}, \label{eq:uomega}
\end{align} 
where we have used the Fourier transformation,
\begin{align}
u(t)=\int \frac{d\omega}{2\pi}u(\omega)e^{-i\omega t}.
\end{align}
In the expression (\ref{eq:uomega}), $l$ is the angular momentum (see also  Appendix \ref{sec:harmonics}) and
\begin{align}
D&=\frac{\eta}{\ebar +\pbar} \label{eq:diffusionconst}
\end{align}
is  the diffusion constant. Therefore, the linear response of the stress tensor and current under the external perturbations are
\begin{align}
\delta T^{ij}(\omega, x^i)&=\sum_I\bigg[-\pbar b_I(\omega)Y_I^{ij}(x^i) +i\omega \eta b_I(\omega) Y_I^{ij}(x^i) \notag \\
&+i \omega\eta \frac{-b_I^0(\omega)  +\frac{\nbar}{\ebar +\pbar}a_I(\omega)}{i\omega -D \frac{(l+8)(l-1)}{R^2}}(\nablabar^i Y_I^j(x^i) +\nablabar^j Y_I^i(x^i))\bigg],  \label{eq:deltaTij} \\ 
\delta T^{0i}(\omega, x^i)&=\sum_I\bigg[\bigg(\ebar+\eta \frac{ \frac{(l+8)(l-1)}{R^2}}{i\omega -D \frac{(l+8)(l-1)}{R^2}}\bigg)b_I^0(\omega)Y_I^i(x^i) \notag \\
&-\nbar \frac{i\omega}{i\omega -D\frac{(l+8)(l-1)}{R^2}}a_I(\omega)Y_I^i(x^i)\bigg], \label{eq:deltaT0i} \\
\delta J^i(\omega, x^i)&=\sum_I\bigg[\bigg(i\omega \sigma-\frac{\nbar^2}{\ebar + \pbar}\frac{i\omega}{i\omega -D\frac{(l+8)(l-1)}{R^2}}\bigg) a_I(\omega)Y_I^i(x^i)  \notag \\
&+\nbar \frac{i\omega}{i\omega -D\frac{(l+8)(l-1)}{R^2}}b_I^0(\omega) Y_I^i(x^i)\bigg]. \label{eq:deltaJi} 
\end{align}
Since we are interested in the dissipative behavior of the stress tensor and current, we neglect the nondissipative terms in (\ref{eq:deltaTij}) and (\ref{eq:deltaT0i}). 
If we expand the stress tensor and current in terms of the spherical harmonics as 
\begin{align}
\delta T^{ij}(\omega, x^i)&=\sum_I T^I(\omega)Y_I^{ij}(x^i)+\tilde{T}^I(\omega)(\nablabar^i Y_I^j(x^i) +\nablabar^j Y_I^i(x^i)), \\
\delta T^{0i}(\omega, x^i)&=\sum_I T^0_I(\omega)Y_I^i(x^i), \\
\delta J^i(\omega,x^i)&=\sum_I J^I(\omega)Y_I^i(x^i),
\end{align}
the coefficients $T^I$, $\tilde T^I$, $T^0_I$ and $J^I$ are 
\begin{align}
T^I(\omega)&=i\omega  \eta b^I(\omega), \label{eq:hydrotensor} \\
\tilde{T}^I(\omega)&=i \omega \eta \frac{-b_I^0(\omega)  +\frac{\nbar}{\ebar +\pbar}a_I(\omega)}{i\omega -D \frac{(l+8)(l-1)}{R^2}}, \\
T_I^0(\omega)&=\eta \frac{ \frac{(l+8)(l-1)}{R^2}}{i\omega -D \frac{(l+8)(l-1)}{R^2}}b_I^0(\omega)-\nbar \frac{i\omega}{i\omega -D\frac{(l+8)(l-1)}{R^2}}a^I(\omega), \label{eq:hydrot0} \\
J^I(\omega)&=\bigg(i\omega \sigma-\frac{\nbar^2}{\ebar + \pbar}\frac{i\omega}{i\omega -D\frac{(l+8)(l-1)}{R^2}}\bigg) a^I(\omega)+\nbar \frac{i\omega}{i\omega -D\frac{(l+8)(l-1)}{R^2}}b_I^0(\omega). \label{eq:hydroj}
\end{align}

\section{Gauge/gravity correspondence for Matrix theory} 
In this section, we briefly review Matrix theory and the gauge/gravity correspondence for Matrix theory in the extremal and the near-extremal case. 
\subsection{Matrix theory} \label{sec:matrix}
Let us consider a system which is composed of $N$ D0-branes on top of one another in 10-dimensional  type IIA string theory.  In this system, there are open strings whose ends are attached on the D0-branes and closed strings which are propagating in the bulk. Although the closed strings are usually coupled to the D0-branes, we can decouple the closed strings  from the D0-branes by taking a near-horizon  limit \cite{Itzhaki:1998dd}.  Since all the massive string modes are also  decoupled in this limit,  the dynamics of the D0-branes can be described by the lowest modes of the open strings, namely, Matrix theory \cite{Banks:1996vh}.

Matrix theory is the maximally supersymmetric $U(N)$ Yang-Mills theory in (0+1) dimensions, which can be viewed as matrix quantum mechanics. The  action is
\begin{align}
S=\int dt \Tr \bigg[\frac{1}{2g_sl_s}\Xdot^m\Xdot_m+\frac{1}{4g_sl_s^5}[X^m,X^n]^2 +(\text{fermionic terms})\bigg],
\end{align}
where  $g_s$ is the string coupling constant and $l_s$ is the string length. In this action, we have adopted the gauge condition $A=0$. The Yang-Mills coupling constant is $\gym^2=(2\pi)^{-2}g_sl_s^{-3}$, which has  mass dimension 3.
 The fields $X^m~(m=1,\cdots, 9)$, which are $N\times N$ Hermitian matrices, describe the lowest modes of open strings connecting the D0-branes and the  diagonal components represent the positions of the D0-branes in the 9 spatial dimensions.

\subsection{Gauge/gravity correspondence: Extremal case} \label{sec:gaugegravityext}
For $g_s\ll1$ and $g_sN \gg 1$, the D0-brane system  can be treated as a classical solution of type IIA supergravity.  The extremal D0-brane black hole solution in string frame is given by%
\footnote{%
Although the overall sign of $A_0$ is different from that in \cite{Sekino:1999av}, this is just a matter of convention. Let us calculate the total R-R charge in our convention.  We apply a  homogeneous chemical potential $\mu\equiv \delta A_0$ to the system at the asymptotic boundary, which is coupled to the total R-R charge $q\equiv \int_{S^8} d^8x \sqrt{g_8}J^0$.  The change of the Lagrangian is $\delta L=\mu q$. The variation of the action with respect to $A_0$ is 
\begin{align}
\delta S_{IIA}&=-\frac{g_s^2}{16\pi G}\int d^{10}x \sqrt{-g}[\nabla_{\mu}(\delta A_{0} F^{\mu 0})+(e.o.m.)] \notag \\
&=\int_{r=\infty} dt \frac{7g_s V_8}{16\pi G R}\delta A_0, \notag
\end{align}
where we have inserted the solution (\ref{eq:stringmetric}) and (\ref{eq:a0back}) in the second line. Therefore, the total R-R charge is $q=+\frac{7g_s V_8}{16\pi G R}$. 
}
\begin{align}
ds_s^2&=-\bigg(1+\frac{R^7}{r^7}\bigg)^{-\frac{1}{2}}dt^2+\bigg(1+\frac{R^7}{r^7}\bigg)^{\frac{1}{2}}(dr^2+r^2 d\Omega_8^2),  \label{eq:stringmetric}\\
e^{\phi}&=g_se^{\phit}=g_s\bigg(1+\frac{R^7}{r^7}\bigg)^{\frac{3}{4}}, \\
A_0&=g_s^{-1}\bigg[\bigg(1+\frac{R^7}{r^7}\bigg)^{-1}-1\bigg], \label{eq:a0back}
\end{align}
where $\phi$ and $A_\mu$ are the dilaton and R-R 1-form field, respectively. 
The ``radius'' $R$ is determined by the number of D0-branes as 
\begin{align}
R&=(60\pi^3)^{\frac{1}{7}} (g_s N)^{\frac{1}{7}} l_s.
\end{align}
By taking the near-horizon limit $R^7/r^7 \gg 1$, the solution becomes
\begin{align}
ds_s^2&=-\bigg(\frac{r}{R}\bigg)^{\frac{7}{2}}dt^2+\bigg(\frac{R}{r}\bigg)^{\frac{7}{2}}(dr^2+r^2 d\Omega_8^2),  \label{eq:extremalsol1} \\
e^{\phit}&=\bigg(\frac{R}{r}\bigg)^{\frac{21}{4}}, \label{eq:extremalsol2} \\
A_0&=\frac{1}{g_s}\frac{r^7}{R^7}. \label{eq:extremalsol3}
\end{align}

For the classical supergravity description to be reliable, the string coupling $e^{\phi}$ must be much smaller than 1 and the curvature radius in string frame must be much longer than $l_s$. 
Then, one finds the following condition for $r$
 \cite{Itzhaki:1998dd,Sekino:1999av}: 
\begin{align}
(g_s N)^{\frac{1}{3}}N^{-\frac{4}{21}} \ll \frac{r}{l_s} \ll (g_s N)^{\frac{1}{3}}. \label{eq:sugracondition}
\end{align}
The former condition leads to the first inequality and the latter condition leads to the second inequality. In addition, from the near-horizon condition $r \ll R$, we have
\begin{align}
\frac{r}{l_s} \ll (g_s N)^{\frac{1}{7}}.
\end{align}
Therefore, the total region of $r$ becomes \cite{Sekino:1999av}
\begin{align}
g_s^{\frac{4}{21}} (g_sN)^{\frac{1}{7}} \ll \frac{r}{l_s} \ll (g_s N)^{\frac{1}{7}}. \label{eq:sugracondition2}
\end{align}
The condition (\ref{eq:sugracondition2}) is satisfied in a wide range of $r$ if $g_s \ll 1$ and $g_sN \gg 1$. Thus,  Matrix theory can be  described by the classical solution (\ref{eq:extremalsol1})--(\ref{eq:extremalsol3}) in this region. 

The near horizon-metric (\ref{eq:extremalsol1}) is related to the metric on $AdS_2\times S^8$  by a Weyl transformation as \cite{Sekino:1999av}
\begin{align}
ds_s^2&=e^{\frac{2}{7}\phit}ds_w^2, \\
ds_w^2&=R^2\bigg[\bigg(\frac{2}{5}\bigg)^2\frac{1}{z^2}(-dt^2+dz^2)+d\Omega_8^2\bigg], \label{eq:extWeyl}
\end{align}
where
\begin{align}
z\equiv \frac{2}{5}R^{\frac{7}{2}}r^{-\frac{5}{2}}.
\end{align}
In our paper, we call the frame whose metric is given by (\ref{eq:extWeyl}) ``AdS frame." 

The GKPW relation for this gauge/gravity correspondence is given by \cite{Sekino:1999av}
\begin{align}
e^{iS_{IIA}[h]}|_{h_I^s(z=z_c)=\bar{h}_I^s}=\bigg\langle \exp \bigg(i \int dt~ \bar{h}_I^s(t) \Ocal_I^s(t) \bigg)\bigg\rangle, \label{eq:GKPW}
\end{align}
where $S_{IIA}$ is the action of 10-dimensional type IIA supergravity, $h_I^s$  denotes  each mode of  perturbations of the  bulk fields, and $z_c$ is the radial coordinate of the cutoff surface, on which Matrix theory is defined.
 On the left-hand side, $h_I^s$ is a solution of  the bulk equations of motion which satisfies two boundary conditions $h_I^s(z\to \infty)=0$  and $h_I^s(z=z_c)=\bar{h}_I^s(t)$. Therefore, the left-hand side is a functional of $\bar{h}_I^s(t)$. The right-hand side represents a generating functional for connected correlation functions of Matrix theory operator $\Ocal_I^s(t)$, which couples to the source $\bar{h}_I^s(t)$. By taking the functional derivatives of  (\ref{eq:GKPW}) with respect to $\bar{h}_I^s(t)$ and sending $\bar{h}_I^s(t)$ to zero, we obtain the correlation functions of Matrix theory via the on-shell action of type IIA supergravity.

\subsection{Gauge/gravity correspondence: Near-extremal case} \label{sec:nearextcorr}
Next, we consider the near-extremal D0-branes. 
If nonextremality is sufficiently small, 
the horizon remains in the near-horizon region $r\ll R$ and 
we can take the near-horizon limit in a similar fashion to the extremal case. 
Then, in the near-horizon limit, the near-extremal D0-brane black hole solution becomes
\begin{align}
ds_s^2&=e^{\frac{2}{7}\phit}ds_w^2, \label{eq:nearext1} \\
ds_w^2&=\Rt^2\bigg[ z^{-2}(-fdt^2+f^{-1}dz^2)+\bigg(\frac{5}{2}\bigg)^2d\Omega_8^2\bigg], \label{eq:nearext2} \\
f&=1-\bigg(\frac{z}{z_0}\bigg)^{\frac{14}{5}}, \notag  \\
e^{\phit}&=\bigg(\frac{z}{\Rt}\bigg)^{\frac{21}{10}}, \label{eq:nearext4} \\
A_0&=\frac{1}{g_s}\bigg(\frac{\Rt}{z}\bigg)^{\frac{14}{5}}, \label{eq:nearext5}
\end{align}
where $\Rt\equiv \frac{2}{5}R$ and $z_0=\frac{2}{5}R^{\frac{7}{2}}r_0^{-\frac{5}{2}}$ denotes the radius of the horizon.  The Hawking temperature $T_H$, the Bekenstein-Hawking entropy $S_{BH}$ and the total D0-brane charge $q$ are\footnote{The D0-brane charge $q$ has mass dimension $1$ because the 1-form field $A_{\mu}$ is defined to be dimensionless.}
\begin{align}
T_H&=\frac{7}{10 \pi z_0}, \\
S_{BH}&=\frac{V_8}{4G}\bigg(\frac{\Rt}{z_0}\bigg)^{\frac{9}{5}}, \\
q&=\frac{7g_s}{16\pi G R}V_8, \label{eq:totalcharge}
\end{align}
where $V_8$ is the volume of $S^8$ with radius $R$.

For the gauge/gravity correspondence to be valid, $r_0$ should be in the region of  (\ref{eq:sugracondition2}). Then, one finds\footnote{The condition (\ref{eq:nearexcondi}) does not hold if we take the large $N$ limit when other parameters are fixed because $\frac{g_{YM}^2N}{T_H^3}\sim (g_s N)^{\frac{5}{2}}(\frac{l_s}{r_0})^{\frac{15}{2}}$.  It is satisfied if $g_s$ scales as $N^{\alpha}~(-1<\alpha < -\frac{3}{7})$ \cite{Sekino:2008he}.}
\begin{align}
(g_sN)^{\frac{10}{7}} \ll \frac{g_{YM}^2N}{T_H^3} \ll N^{\frac{10}{7}}. \label{eq:nearexcondi}
\end{align}
Thus, we can study the strongly coupled D0-brane system by using the gauge/gravity correspondence.

In the near-extremal case,  the definition of the GKPW relation (\ref{eq:GKPW}) is subtle if we discuss the real time correlation functions \cite{Herzog:2002pc}. 
Since the regularity condition on the horizon is not well-defined in this case, 
we impose the ingoing boundary condition 
\cite{Policastro:2002se,Son:2002sd}. 
Up to the quadratic order of the perturbations, the on-shell action takes the following form:
\begin{align}
S_{\mathrm{on-shell}}&=\sum_I A_I\int \frac{d\omega}{2\pi} \bar{h}_I^s(-\omega) \Fcal_I(\omega, z) \bar{h}_I^{s'}(\omega)\big|_{z=z_c},
\end{align}
where 
\begin{align}
A_I=
\begin{cases}
D_2^I\equiv \frac{1}{2}\int d^8x \sqrt{g_8}Y_I^{ij}Y^I_{ij}, ~~~~~~~(\text{for tensor mode}), \\
D_1^I\equiv \int d^8x \sqrt{g_8}Y_I^iY_i^I, ~~~~~~~~~~(\text{for vector modes}),
\end{cases}
\end{align}
and $\sqrt{g_8}$ is the square root of the determinant of the metric on $S^8$ with radius $R$. 
Hereafter, we suppress the angular momentum index $I$ and the sum over $I$. Then, the retarded Green function of Matrix theory is given by
\begin{align}
G^{ss'}_R(\omega)=
\begin{cases}
-2 \Fcal(\omega, z)|_{z= z_c},~~~~~~~~(\text{for}~s=s'), \\
- \Fcal(\omega, z)|_{z= z_c},~~~~~~~~~~(\text{for}~s\neq s'), 
\end{cases} \label{eq:prescription}
\end{align}
where the retarded Green function of operators $\mathcal O^s$ is defined by 
\begin{align}
G^{ss'}_R(\omega)&=-i\int_{-\infty}^{\infty}dt e^{i\omega t} \theta(t) \langle [\Ocal^s(t), \Ocal^{s'}(0)] \rangle_0. 
\end{align}
Here, $\langle ~~ \rangle_0$ denotes the ensemble average in equilibrium.
Thus, according to the linear response theory \cite{Chaikin:book}, the linear response of the Matrix theory operator is 
\begin{align}
\delta \langle \Ocal^s(\omega) \rangle&\equiv \langle \Ocal^s(\omega) \rangle -\langle \Ocal^s(\omega) \rangle_0 \notag \\
&=-G_R^{ss'}(\omega)\bar{h}^{s'}(\omega), 
\end{align}
where  $\langle ~~ \rangle$ denotes the ensemble average when the source fields $\bar{h}^s(\omega)$ are turned on.

\section{Linear responses of D0-branes} \label{sec:linearresponse}
In this section, we investigate fluid in the gauge/gravity correspondence for Matrix theory. 
We introduce perturbations in tensor and vector modes and 
calculate the linear responses of the stress tensor and R-R 1-form current. 
Our strategy for the calculation is the following:
\begin{itemize}
\item We put the cutoff surface at $z=z_c$.
\item We solve the bulk equations of motion for the perturbations of the metric and the R-R 1-form. Then, we impose the ingoing boundary condition at $z=z_0$ and the Dirichlet boundary condition at $z=z_c$ on the solutions.
\item We evaluate the on-shell action and calculate the linear responses of the operators which are coupled to those perturbations. 
\item We compare the results with the hydrodynamic stress tensor and current on $S^8$ (or $\mathbf{R}^8$) when 
\begin{itemize}
\item[(a)] the cutoff surface is near the horizon. 
\item[(b)] the cutoff surface is far from the horizon.
\end{itemize}
\end{itemize}

The bosonic action of the 10-dimensional type IIA supergravity in string frame is%
\footnote{
Hereafter, we denote $\phit$ as $\phi$ for simplicity.}
\begin{align}
S_{IIA}=\frac{1}{2\kappa^2}\int d^{10}x \sqrt{-g}\bigg[e^{-2\phi}(R+4\del_{\mu}\phi \del^{\mu}\phi )-\frac{g_s^2}{4}F_{\mu\nu}F^{\mu\nu}\bigg],
\end{align}
where $\mu, \nu=0,\cdots, 9$; $2\kappa^2=16\pi G$; and $G$  is the 10-dimensional Newton constant. 
Because Neveu-Schwarz--Neveu-Schwarz 2-form and R-R 3-form have no nontrivial backgrounds in the D0-brane solution (\ref{eq:nearext1})--(\ref{eq:nearext5}), 
they are decoupled from the metric and R-R 1-form at the linear order. 
Hence, we have omitted them.

To reduce the calculations, it is convenient to use the AdS frame (\ref{eq:nearext2}). 
Then, the action becomes \cite{Sekino:1999av}
\begin{align}
S_{IIA}'=\frac{1}{2\kappa^2}\int d^{10}x \sqrt{-g}e^{-\frac{6}{7}\phi}\bigg(R+\frac{16}{49}\del_{\mu}\phi \del^{\mu}\phi -\frac{g_s^2}{4}e^{\frac{12}{7}\phi}F_{\mu\nu}F^{\mu\nu}\bigg). \label{eq:IIA'}
\end{align}
To obtain the correct on-shell action, we need to add the  Gibbons-Hawking term \cite{Gibbons:1976ue},
\begin{align}
S_{GH}&=-\frac{1}{\kappa^2}\int_{z=z_c}d^9x \sqrt{-\gamma}e^{-\frac{6}{7}\phi}K, \label{eq:GB}
\end{align}
on the boundary, where  $\gamma_{\mu\nu}$ is the induced metric on the cutoff surface and $K$ is the trace of the extrinsic curvature $K_{\mu\nu}$. 
Therefore, the total action is 
\begin{align}
S_{total}&=S_{IIA}'+S_{GH}. \label{eq:total}
\end{align}

By varying the action (\ref{eq:IIA'}) with respect to the metric, 1-form and dilaton, one finds the equations of motion
\begin{align}
0&=R_{\mu\nu}-\frac{1}{2}g_{\mu\nu}R+\frac{4}{7}(g_{\mu\nu}\del_{\rho}\phi \del^{\rho}\phi-\frac{5}{7}\del_{\mu}\phi \del_{\nu}\phi)+\frac{6}{7}(\nabla_{\mu}\del_{\nu}\phi-g_{\mu\nu}\nabla_{\rho}\del^{\rho}\phi) \notag \\
&+\frac{g_s^2}{8}e^{\frac{12}{7}\phi}(g_{\mu\nu}F_{\rho\sigma}F^{\rho\sigma}-4F_{\mu\rho}F_{\nu}{}^{\rho}), \label{eq:eommetric} \\
0&=\nabla_{\mu}(e^{\frac{6}{7}\phi}F^{\mu\nu}), \label{eq:eom1form} \\
0&=R-\frac{16}{49}\del_{\mu}\phi\del^{\mu}\phi+\frac{16}{21}\nabla_{\mu}\nabla^{\mu}\phi+\frac{g_s^2}{4}e^{\frac{12}{7}\phi}F_{\mu\nu}F^{\mu\nu}.
\end{align}
We denote the linear perturbations of the metric, R-R 1-form and dilaton around the background fields (\ref{eq:nearext2})-(\ref{eq:nearext5}) as $h_{\mu\nu}, \Ahat_{\mu}$, and $\phihat$, respectively. 
At the linear order of the perturbations,  the equations of motion become
\begin{align}
0&=\nabla_{\mu}\del_{\nu}h+(\nabla^{\rho}\nabla_{\rho}h_{\mu\nu} -\nabla_{\rho}\nabla_{\mu}h^{\rho}{}_{\nu}-\nabla_{\rho}\nabla_{\nu}h^{\rho}{}_{\mu}) \notag \\
&+h_{\mu\nu}R-g_{\mu\nu}(h^{\rho}{}_{\sigma}R^{\sigma}{}_{\rho}+\nabla^{\rho}\del_{\rho}h-\nabla_{\rho}\nabla^{\sigma}h^{\rho}{}_{\sigma}) \notag \\
&-\frac{8}{7}\bigg[h_{\mu\nu}\del_{\rho}\phi\del^{\rho}\phi-g_{\mu\nu}(h^{\rho}{}_{\sigma}\del_{\rho}\phi\del^{\sigma}\phi -2\del_{\rho}\phihat\del^{\rho}\phi)-\frac{5}{7}(\del_{\mu}\phihat\del_{\nu}\phi+\del_{\mu} \phi\del_{\nu}\phihat)\bigg] \notag \\
&-\frac{12}{7}\bigg[\nabla_{\mu}\del_{\nu}\phihat-g_{\mu\nu}\nabla_{\rho}\del^{\rho}\phihat-\frac{1}{2}(\nabla_{\nu}h^{\rho}{}_{\mu}+\nabla_{\mu}h^{\rho}{}_{\nu}-\nabla^{\rho}h_{\mu\nu})\del_{\rho}\phi \notag \\
&-h_{\mu\nu}\nabla_{\rho}\del^{\rho}\phi+g_{\mu\nu}(\nabla_{\rho}h^{\rho}{}_{\sigma}\del^{\sigma}\phi+h^{\rho}{}_{\sigma}\nabla_{\rho}\del^{\sigma}\phi-\frac{1}{2}\del_{\rho}h\del^{\rho}\phi)\bigg] \notag \\
&-\frac{g_s^2}{4}e^{\frac{12}{7}\phi}\bigg[h_{\mu\nu}F_{\rho\sigma}F^{\rho\sigma}+4h^{\rho}{}_{\sigma}F_{\mu\rho}F_{\nu}{}^{\sigma}-2g_{\mu\nu}h^{\rho}{}_{\lambda}F_{\rho\sigma}F^{\lambda\sigma} \notag \\
&+2g_{\mu\nu}\Fhat_{\rho\sigma}F^{\rho\sigma}-4(\Fhat_{\mu\rho}F_{\nu}{}^{\rho}+F_{\mu\rho}\Fhat_{\nu}{}^{\rho})+\frac{12}{7}\phihat (g_{\mu\nu}F_{\rho\sigma}F^{\rho\sigma}-4F_{\mu\rho}F_{\nu}{}^{\rho})\bigg], \label{eq:metricpert} \\
0&=\frac{6}{7}(\del_{\nu}\phihat F^{\nu\mu}+\del_{\nu}\phi \Fhat^{\nu\mu}-\del_{\nu}\phi h^{\nu}{}_{\rho}F^{\rho\mu})+\nabla_{\nu}\Fhat^{\nu\mu} \notag \\
&+\frac{1}{2}\del_{\nu}hF^{\nu\mu}-\nabla_{\nu}h^{\nu}{}_{\rho}F^{\rho\mu}-h^{\nu}{}_{\rho}\nabla_{\nu}F^{\rho\mu}-\nabla_{\nu}h^{\mu}{}_{\rho}F^{\nu\rho}, \label{eq:1formpert} \\
0&=-\nabla_{\mu}\del^{\mu}h+\nabla_{\mu}\nabla^{\nu}h^{\mu}{}_{\nu}-h^{\mu}{}_{\nu}R^{\nu}{}_{\mu}-\frac{16}{49}(2\del_{\mu}\phihat\del^{\mu}\phi-h^{\mu}{}_{\nu}\del_{\mu}\phi\del^{\nu}\phi) \notag \\
&+\frac{16}{21}\bigg(\frac{1}{2}\del_{\mu}h\del^{\mu}\phi-\nabla_{\mu}h^{\mu}{}_{\nu}\del^{\nu}\phi-h^{\mu}{}_{\nu}\nabla_{\mu}\del^{\nu}\phi+\nabla_{\mu}\del^{\mu}\phihat\bigg) \notag \\
&+\frac{g_s^2}{2}e^{\frac{12}{7}\phi}\bigg(\frac{6}{7}\phihat F_{\mu\nu}F^{\mu\nu}-h^{\mu}{}_{\rho}F_{\mu\nu}F^{\rho\nu}+\Fhat_{\mu\nu}F^{\mu\nu}\bigg),  \label{eq:dilatonpert}
\end{align}
where $\Fhat_{\mu\nu}=\del_{\mu}\Ahat_{\nu}-\del_{\nu}\Ahat_{\mu}$.

We adopt the following gauge conditions \cite{Sekino:1999av}:
\begin{align}
&\nabla_i \big(h^i{}_j-\frac{1}{8}\delta^i_j h^k{}_k\big)=0, \\
&\nabla^i h_i^0=\nabla^ih_i^z=0, \\
&\nabla^i \Ahat_i=0,
\end{align}
where $i=1,\cdots, 8$.
Then, using the spherical harmonic expansions on $S^8$, 
we can classify  the perturbations into the scalar modes,
\begin{align}
h_0^0(x^{\mu})&=\sum b^0_0(t,z) Y(x^i),~~~~h_z^0(x^{\mu})=\sum b^0_z(t,z) Y(x^i), \notag \\
h_z^z(x^{\mu})&=\sum b^z_z(t,z) Y(x^i),~~~~h_i^i(x^{\mu})=\sum b^i_i(t,z) Y(x^i), \notag \\
\Ahat_0(x^{\mu})&=\sum a_0(t,z)Y(x^i),~~~~\Ahat_z(x^{\mu})=\sum a_z(t,z) Y(x^i), \notag \\
\phihat(x^{\mu})&=\sum \varphi(t,z) Y(x^i), \label{eq:scalarmode}
\end{align}
vector modes,
\begin{align}
h_i^0(x^{\mu})&=\sum b^0(t,z) Y_i(x^i),~~~~h_i^z(x^{\mu})=\sum b^z(t,z) Y_i(x^i), \notag \\
\Ahat_i(x^{\mu})&=\sum a(t,z) Y_i(x^i), \label{eq:vectormode}
\end{align}
and  tensor mode,
\begin{align}
h^i{}_j(x^{\mu})-\frac{1}{8}\delta^i{}_jh^k{}_k(x^{\mu})=\sum b(t,z)Y^i{}_{j}(x^i), \label{eq:tensormode}
\end{align}
where $Y, Y_i$, and $Y_{ij}$ are the scalar, vector, and tensor harmonics on $S^8$, respectively. 
Here, we have suppressed the angular momentum indices. 
Since these modes are decoupled from each other, 
we can analyze each mode independently.

\subsection{Solutions of equations of motion}  \label{sec:sol}
As we will see later, 
the equations of motion can be reduced into the differential equations 
which generally take the following form:
\begin{align}
0&=f^{-1}u^{-p}(u^pf\chi')'+\omegat^2f^{-2}u^{-\frac{9}{7}}\chi-k^2f^{-1}u^{-2}\chi \notag \\
&=\chi''+\bigg(\frac{p}{u}-\frac{1}{1-u}\bigg)\chi'+\frac{\omegat^2}{u^{\frac{9}{7}}(1-u)^2}\chi-\frac{k^2}{u^2(1-u)}\chi. \label{eq:generaleq}
\end{align}
In this section, we will discuss the solutions and boundary conditions 
for this differential equation. 
The function $\chi$ 
is related to perturbations of the metric and R-R 1-form. 
The parameters $p$ and $k$ take 
\begin{align}
p=0,~~~~~k^2=\frac{l(l+7)}{49},
\end{align}
for the tensor mode, 
and 
\begin{align}
p&=\frac{9}{7},~~~~~k^2=\frac{(l+1)(l-1)}{49},\frac{(l+6)(l+8)}{49}, 
\end{align}
for two independent modes of the vector modes. 
The variable $u$ is related to the radial coordinate $z$ as 
\begin{align}
u=\bigg(\frac{z}{z_0}\bigg)^{\frac{14}{5}}, \label{eq:defu}
\end{align}
and $\omegat$ is the dimensionless frequency,
\begin{equation}
 \omegat = \frac{\omega}{4\pi T_H} . 
\end{equation}
Although it is  difficult to solve  the differential equation (\ref{eq:generaleq}) for an arbitrary $\omegat$, we can obtain the solution in the hydrodynamic regime, $\omegat \ll 1$.

Near the horizon $u\simeq 1$, the leading contributions of 
the differential equation (\ref{eq:generaleq}) are
\begin{align}
0=\chi''-\frac{1}{1-u}\chi'+\frac{\omegat^2}{(1-u)^2}\chi. \label{eq:nearheq}
\end{align}
Then, the leading terms of two independent solutions of \eqref{eq:generaleq} are 
\begin{align}
\chi(\omegat, u) &=C_1(1-u)^{-i\omegat}+C_2(1-u)^{i\omegat} \notag \\
&\stackrel{\omegat \ll 1}{\simeq}C_1(1-i\omegat \ln(1-u))+C_2(1+i\omegat \ln(1-u)), \label{eq:ingoingcond}
\end{align}
where $C_1$ and  $C_2$ are the integration constants.
Imposing the ingoing boundary condition,  $C_2$ must vanish \cite{Policastro:2002se,Son:2002sd}.

For $\omegat \ll 1$,  $\chi$ can be expanded as a series of $\omegat^2$ as 
\begin{align}
\chi(\omegat,u)=\chi_0(u)+\omegat^2\chi_2(u) +\omegat^4\chi_4(u)+\cdots, 
\end{align}
and the coefficients $\chi_n(u)$ can be solved recursively. 
The differential equation for $\chi_0$ is
\begin{align}
0=\chi_0''+\bigg(\frac{p}{u}-\frac{1}{1-u}\bigg)\chi_0'-\frac{k^2}{u^2(1-u)}\chi_0 , \label{eq:x0eq}
\end{align}
and the solution is\footnote{When $\alpha+\beta \in \mathbf{Z}~ (l\in 7\mathbf{Z})$, the solution of (\ref{eq:x0eq}) is not given by (\ref{eq:solchi0}) \cite{Abramowitz:book}. In this paper, we do not deal with this exceptional case.}
\begin{align}
\chi_0=\Ctilde_1u^{\alpha}{}_2F_1(\alpha, \beta; \alpha+\beta;u)+\Ctilde_2 u^{\gamma}{}_2F_1(\gamma, \delta; \gamma+\delta;u), \label{eq:solchi0}
\end{align}
where $\Ctilde_1$ and  $\Ctilde_2$ are the integration constants, ${}_2F_1$ is the Gauss' hypergeometric function and 
\begin{align}
\alpha&=\frac{1}{2}(1-p-\sqrt{(1-p)^2+4k^2}), \\
\beta&=\frac{1}{2}(1+p-\sqrt{(1-p)^2+4k^2}), \\
\gamma&=\frac{1}{2}(1-p+\sqrt{(1-p)^2+4k^2}), \\
\delta&=\frac{1}{2}(1+p+\sqrt{(1-p)^2+4k^2}).
\end{align}

The  expansion of  the Gauss' hypergeometric function  ${}_2F_1(a,b;c;u)$ around $u=1$ is special when $a+b=c$. It is given by \cite{Abramowitz:book} 
\begin{align}
&{}_2F_1(a,b;a+b;u)= \notag \\
&\frac{\Gamma(a+b)}{\Gamma(a)\Gamma(b)}\sum_{n=0}^{\infty} \frac{(a)_n(b)_n}{(n!)^2}[2\psi (n+1)-\psi (a+n)-\psi(b+n)-\ln (1-u)](1-u)^n,
\end{align}
where  $\psi(n)$ is the digamma function and
\begin{align}
(a)_n&=a(a+1)(a+2)\cdots (a+n-1), ~~~~(a)_0=1, \\
\psi(1)&=-\gamma_e=-0.57721\cdots,~~~~~~~(\gamma_e: \text{Euler constant}).
\end{align}
Thus, near the horizon $u\simeq 1$, $\chi_0$ becomes
\begin{align}
\chi_0&\stackrel{u\to 1}{\simeq}\Ctilde_1\frac{\Gamma(\alpha+\beta)}{\Gamma(\alpha)\Gamma(\beta)}[-2\gamma_e -\psi(\alpha)-\psi(\beta)-\ln (1-u)] \notag \\
 &+\Ctilde_2\frac{\Gamma(\gamma+\delta)}{\Gamma(\gamma)\Gamma(\delta)}[-2\gamma_e -\psi(\gamma)-\psi(\delta)-\ln (1-u)].
\end{align}
This can be summarized in the form of (\ref{eq:ingoingcond}).  Since $C_2=0$ from the ingoing boundary condition, we find
\begin{align}
\Ctilde_2&=-\Ctilde_1\frac{\Gamma(\alpha+\beta)}{\Gamma(\alpha)\Gamma(\beta)}\frac{\Gamma(\gamma)\Gamma(\delta)}{\Gamma(\gamma+\delta)}\frac{1+i\omegat(2\gamma_e +\psi(\alpha)+\psi(\beta))}{1+i\omegat(2\gamma_e +\psi(\gamma)+\psi(\delta))} \notag \\
&\stackrel{\omegat \to 0}{\simeq}-\Ctilde_1\frac{\Gamma(\alpha+\beta)}{\Gamma(\alpha)\Gamma(\beta)}\frac{\Gamma(\gamma)\Gamma(\delta)}{\Gamma(\gamma+\delta)}(1+i\omegat (\psi(\alpha)+\psi(\beta)-\psi(\gamma)-\psi(\delta))).
\end{align}
Imposing the Dirichlet boundary condition at the cutoff surface,
\begin{align}
\chi(\omegat, u_c)=\bar{\chi}(\omegat),
\end{align}
the solution is
\begin{align}
\chi(\omegat, u)&=\frac{\bar{\chi}(\omegat)}{\Fscr}\bigg[u^{\alpha}{}_2F_1(\alpha,\beta;\alpha+\beta;u) \notag \\
&-\frac{\Gamma(\alpha+\beta)}{\Gamma(\alpha)\Gamma(\beta)}\frac{\Gamma(\gamma)\Gamma(\delta)}{\Gamma(\gamma+\delta)}(1+i\omegat (\psi(\alpha)+\psi(\beta)-\psi(\gamma)-\psi(\delta)))u^{\gamma}{}_2F_1(\gamma,\delta;\gamma+\delta;u)\bigg] \notag \\
&+\Ocal(\omegat^2),
\end{align}
where 
\begin{align}
\Fscr&=u_c^{\alpha}{}_2F_1(\alpha,\beta;\alpha+\beta;u_c) \notag \\ &-\frac{\Gamma(\alpha+\beta)}{\Gamma(\alpha)\Gamma(\beta)}\frac{\Gamma(\gamma)\Gamma(\delta)}{\Gamma(\gamma+\delta)}(1+i\omegat (\psi(\alpha)+\psi(\beta)-\psi(\gamma)-\psi(\delta)))u_c^{\gamma}{}_2F_1(\gamma,\delta;\gamma+\delta;u_c).
\end{align}

\subsection{Tensor mode} \label{sec:tensor}
Now, we solve the equations of motion, and calculate the on-shell action. 
We first consider the tensor mode. 
The equation of motion for the tensor mode $b$ is
\begin{align}
0=f^{-1}z^{\frac{9}{5}}\del_z(fz^{-\frac{9}{5}}\del_z b)-f^{-2}\del_0^2 b-\frac{4}{25}l(l+7)f^{-1}z^{-2}b. \label{eq:eomb}
\end{align}
For the tensor mode, the angular momentum must satisfy $l\geq 2$. 
Using the Fourier transformation,
\begin{align}
b(t,u)&=\int \frac{d\omega}{2\pi} b(\omega,u) e^{-i\omega t},
\end{align}
and the coordinate $u$ which is defined by \eqref{eq:defu}, 
the equation of motion  (\ref{eq:eomb}) becomes
\begin{align}
0=f^{-1}(fb')'+\omegat^2 u^{-\frac{9}{7}}f^{-2}b-\frac{l(l+7)}{49}u^{-2}f^{-1}b, \label{eq:bdiffeq}
\end{align}
where $\omegat$ is the dimensionless frequency, 
\begin{align}
\omegat=\frac{5z_0}{14}\omega=\frac{\omega}{4\pi T_H}.
\end{align}
The prime $'$ denotes the derivative with respect to $u$.
From the result of  section \ref{sec:sol}, 
the solution of  (\ref{eq:bdiffeq}) for $\omegat \ll 1$  is
\begin{align}
b(\omegat, u)=\bar{b}(\omegat)\frac{u^{-\frac{l}{7}}F(u) -X(\omegat)u^{1+\frac{l}{7}}\Ftilde(u)}{u_c^{-\frac{l}{7}}F(u_c) -X(\omegat)u_c^{1+\frac{l}{7}}\Ftilde(u_c)}, \label{eq:solutionofb}
\end{align}
where  $u_c=(z_c/z_0)^{{14}/{5}}$ and
\begin{align}
\bbar(\omegat)&\equiv b(\omegat, u_c), \\
F(u)&\equiv {}_2F_1\bigg(\!-\frac{l}{7},-\frac{l}{7};-\frac{2l}{7};u\bigg),~~~~~\Ftilde(u)\equiv {}_2F_1\bigg(1+\frac{l}{7},1+\frac{l}{7};2+\frac{2l}{7};u\bigg), \\
X(\omegat)&\equiv \frac{\Gamma(-\frac{2l}{7})}{\Gamma(-\frac{l}{7})^2}\frac{\Gamma(1+\frac{l}{7})^2}{\Gamma(2+\frac{2l}{7})} \bigg[1+2\pi i \omegat \cot \bigg(\frac{l\pi}{7}\bigg)\bigg].
\end{align}

The on-shell action for $b$ is
\begin{align}
2\kappa^2S_{\mathrm{on-shell}}&=\frac{7}{5}\Rt^{\frac{9}{5}}z_0^{-\frac{14}{5}}D_2\int_{u=u_c}dt [fbb'-u^{-1}(1+f)b^2]. \label{eq:onshellactionb}
\end{align}
Derivation of the on-shell action is given in  Appendix \ref{sec:tensoronshell}. Hereafter, we do not consider the contact term, which does not contribute to the dissipative behavior. Inserting (\ref{eq:solutionofb}) into the on-shell action, we find
\begin{align}
2\kappa^2S_{\mathrm{on-shell}}&=\frac{7}{5}\Rt^{\frac{9}{5}}z_0^{-\frac{14}{5}}D_2\int \frac{d\omega}{2\pi}(1-u_c)\bbar(-\omega)\bbar(\omega) \notag \\
&\cdot \frac{-\frac{l}{7}u_c^{-1}F(u_c)+F'(u_c)-Xu_c^{\frac{2l}{7}}((1+\frac{l}{7})\Ftilde(u_c)+u_c \Ftilde'(u_c))}{F(u_c) -Xu_c^{1+\frac{2l}{7}}\Ftilde(u_c)}. \label{eq:tensoronshellgen}
\end{align}

\subsection{Vector modes} \label{sec:vector}
Next, we consider the vector modes. 
In this case, the vector modes  consist of metric components  $b^0, b^z$, and R-R 1-form $a$. 
There are four equations of motion for these fields,
\begin{align}
0&=\del_z^2b^0+f^{-2}\del_0\del_zb^z-\bigg(\frac{19}{5}z^{-1}-2f^{-1}\del_zf\bigg)\del_zb^0-\bigg(\frac{9}{5}z^{-1}f^{-2}+f^{-3}\del_zf\bigg)\del_0b^z \notag \\
&-\frac{4}{25}((l+1)(l+6)-49)z^{-2}f^{-1}b^0+\frac{14}{5}g_s\Rt^{-\frac{14}{5}}z^{\frac{9}{5}}f^{-1}\del_za,~~~~~(l\geq 1), \label{eq:eqofb0} \\
0&=-f^{-2}\del_0^2b^z+(2z^{-1}-f^{-1}\del_zf)\del_0b^0-\del_0\del_zb^0 \notag \\
&-\frac{4}{25}(l+8)(l-1)f^{-1}z^{-2}b^z-\frac{14}{5}g_sz^{\frac{9}{5}}\Rt^{-\frac{14}{5}}f^{-1}\del_0a,~~~~~(l\geq 1), \label{eq:eqofbz} \\
0&=\del_z(f\del_za)+\frac{9}{5}z^{-1}f\del_za-f^{-1}\del_0^2a-\frac{4}{25}(l+1)(l+6)z^{-2}a \notag \\
&+\frac{14}{5}g_s^{-1}\Rt^{\frac{14}{5}}z^{-\frac{19}{5}}[\del_z(fb^0)+f^{-1}\del_0b^z-2z^{-1}fb^0],~~~~~(l\geq 1),  \label{eq:eqofa}\\
0&=\del_0b^0+\del_zb^z-\frac{19}{5}z^{-1}b^z,~~~~~(l\geq 2). \label{eq:eqofv} 
\end{align}
The equation (\ref{eq:eqofv}) can be derived from (\ref{eq:eqofb0}) and (\ref{eq:eqofbz}). Since  two equations (\ref{eq:eqofbz}) and (\ref{eq:eqofa})  contain the  second order time derivative, there are two physical degrees of freedom in the vector modes. 
The other equation (\ref{eq:eqofv}), or equivalently (\ref{eq:eqofb0}) 
gives a constraint on the boundary conditions. 
Therefore, these equations of motion yield two second order differential equations and 
one first order equation. 
The solution has five integration constants which can be fixed 
by two incoming boundary conditions at the horizon and 
three Dirichlet boundary conditions on the cutoff surface 
for $b^0, b^z$, and $a$. 

Let us set \cite{Sekino:1999av}
\begin{align}
\ahat&=-g_s a, \\
\bhat&=-\frac{5}{14}\Rt^{\frac{14}{5}}z^{-\frac{9}{5}}(z^2 \del_z(z^{-2}fb^0)+f^{-1}\del_0 b^z). \label{eq:defbhat}
\end{align}
Then, the equations (\ref{eq:eqofb0}), (\ref{eq:eqofbz}) and  (\ref{eq:eqofa}) become
\begin{align}
0&=\del_z \bhat +\del_z \ahat+\frac{2}{35}(l+8)(l-1)\Rt^{\frac{14}{5}}z^{-\frac{19}{5}}b^0, \label{eq:eqofb01} \\
0&=\del_0 \bhat +\del_0 \ahat -\frac{2}{35}(l+8)(l-1)\Rt^{\frac{14}{5}}z^{-\frac{19}{5}}b^z, \label{eq:eqofbz1} \\
0&=z^{\frac{1}{5}}\del_z(z^{\frac{9}{5}}f\del_z \ahat)-z^2f^{-1}\del_0^2 \ahat-\frac{4}{25}(l+1)(l+6)\ahat+\frac{196}{25}\bhat . \label{eq:ahat}
\end{align}
Inserting (\ref{eq:eqofb01}) and (\ref{eq:eqofbz1}) into (\ref{eq:defbhat}), we find
\begin{align}
0=z^{\frac{1}{5}}\del_z (z^{\frac{9}{5}} f \del_z(\ahat+\bhat))-z^2f^{-1}\del_0^2(\ahat+\bhat)-\frac{4}{25}(l+8)(l-1)\bhat. \label{eq:ahatbhat}
\end{align}
Using (\ref{eq:ahat}), the equation (\ref{eq:ahatbhat}) becomes
\begin{align}
0=z^{\frac{1}{5}}\del_z(z^{\frac{9}{5}}f\del_z \bhat)-z^2f^{-1}\del_0^2 \bhat+\frac{4}{25}(l+1)(l+6)\ahat-\frac{4}{25}((l+8)(l-1)+49)\bhat. \label{eq:bhat}
\end{align}
Thus, we have obtained two equations (\ref{eq:ahat}) and (\ref{eq:bhat}) for $\ahat$ and $\bhat$, which represent the two physical degrees of freedom of the vector modes. 

To solve the equations, let us set
\begin{align}
\bbf&=
\begin{pmatrix}
\ahat \\
\bhat
\end{pmatrix}, \\
\Mbf&=
\begin{pmatrix}
(l+1)(l+6) & -49 \\
-(l+1)(l+6) & (l+8)(l-1)+49
\end{pmatrix}.
\end{align}
Then, the equations (\ref{eq:ahat}) and (\ref{eq:bhat}) can be summarized in the following form,
\begin{align}
0&=z^{\frac{1}{5}}\del_z(z^{\frac{9}{5}}f\del_z \bbf)-z^2f^{-1}\del_0^2 \bbf-\frac{4}{25}\Mbf \cdot \bbf. \label{eq:undiag}
\end{align}
Since the eigenmatrix  of $\Mbf$ is
\begin{align}
\Lambdabf&=\Ubf^{-1}\Mbf \Ubf, \notag \\
&=
\begin{pmatrix}
(l+1)(l-1) & 0 \\
0 & (l+6)(l+8)
\end{pmatrix}, 
\end{align}
where
\begin{align}
\Ubf&=
\begin{pmatrix}
1 & 1 \\
\frac{l+1}{7} & -\frac{l+6}{7}
\end{pmatrix},
\end{align}
the equation (\ref{eq:undiag}) can be diagonalized as
\begin{align}
0&=z^{\frac{1}{5}}\del_z(z^{\frac{9}{5}}f\del_z \abf)-z^2f^{-1}\del_0^2 \abf-\frac{4}{25}\Lambdabf \cdot \abf,
\end{align}
where
\begin{align}
\abf\equiv 
\begin{pmatrix}
\ahat_1 \\
\ahat_2
\end{pmatrix}
=\Ubf^{-1} \bbf. \label{eq:aandb}
\end{align}
Therefore, we obtain the diagonalized equations of motion,
\begin{align}
0&=f^{-1}u^{-\frac{9}{7}}(u^{\frac{9}{7}}f\ahat_1')'+f^{-2}u^{-\frac{9}{7}}\omegat^2 \ahat_1-\frac{(l+1)(l-1)}{49}f^{-1}u^{-2}\ahat_1, \\
0&=f^{-1}u^{-\frac{9}{7}}(u^{\frac{9}{7}}f\ahat_2')'+f^{-2}u^{-\frac{9}{7}}\omegat^2 \ahat_2-\frac{(l+6)(l+8)}{49}f^{-1}u^{-2}\ahat_2,
\end{align}
where we have used the Fourier transformations of $\ahat_1$ and  $\ahat_2$.

From (\ref{eq:eqofb01}), (\ref{eq:eqofbz1}), and (\ref{eq:aandb}), the original modes $\ahat, b^0$, and  $b^z$ are 
\begin{align}
\ahat&=\ahat_1+\ahat_2, \label{eq:ahat12} \\
b^0&=-7 \bigg(\frac{z_0}{\Rt}\bigg)^{\frac{14}{5}}u^2\bigg(\frac{\ahat_1'}{l-1}-\frac{\ahat_2'}{l+8}\bigg), \label{eq:b012} \\
b^z&=-7i\omegat \bigg(\frac{z_0}{\Rt}\bigg)^{\frac{14}{5}}u^{\frac{19}{14}}\bigg(\frac{\ahat_1}{l-1}-\frac{\ahat_2}{l+8}\bigg). \label{eq:bza1a2}
\end{align}

From the result of section \ref{sec:sol}, the solutions of the diagonalized equations of motion  for $\omegat \ll 1$ are
\begin{align}
\ahat_1(\omegat, u)&=\abar_1(\omegat)\frac{u^{-\frac{1}{7}-\frac{l}{7}}F_1(u)-X_1(\omegat)u^{-\frac{1}{7}+\frac{l}{7}}\Ftilde_1(u)}{u_c^{-\frac{1}{7}-\frac{l}{7}}F_1(u_c)-X_1(\omegat)u_c^{-\frac{1}{7}+\frac{l}{7}}\Ftilde_1(u_c)}, \\
\ahat_2(\omegat, u)&=\abar_2(\omegat)\frac{u^{-\frac{8}{7}-\frac{l}{7}}F_2(u)-X_2(\omegat)u^{\frac{6}{7}+\frac{l}{7}}\Ftilde_2(u)}{u_c^{-\frac{8}{7}-\frac{l}{7}}F_2(u_c)-X_2(\omegat)u_c^{\frac{6}{7}+\frac{l}{7}}\Ftilde_2(u_c)},
\end{align}
where $\abar_{1,2}(\omegat)\equiv \ahat_{1,2}(\omegat, u_c)$ and
\begin{align}
F_1&\equiv {}_2F_1\bigg(\! -\frac{1}{7}-\frac{l}{7}, \frac{8}{7}-\frac{l}{7};1-\frac{2l}{7};u\bigg),~~\Ftilde_1\equiv {}_2F_1\bigg(\! -\frac{1}{7}+\frac{l}{7}, \frac{8}{7}+\frac{l}{7};1+\frac{2l}{7};u\bigg), \\
F_2&\equiv {}_2F_1\bigg(\! -\frac{8}{7}-\frac{l}{7}, \frac{1}{7}-\frac{l}{7};-1-\frac{2l}{7};u\bigg),~~\Ftilde_2\equiv {}_2F_1\bigg(\frac{6}{7}+\frac{l}{7}, \frac{15}{7}+\frac{l}{7};3+\frac{2l}{7};u\bigg), \\
X_1&\equiv \frac{\Gamma(1-\frac{2l}{7})}{\Gamma(-\frac{1}{7}-\frac{l}{7})\Gamma(\frac{8}{7}-\frac{l}{7})}\frac{\Gamma(-\frac{1}{7}+\frac{l}{7})\Gamma(\frac{8}{7}+\frac{l}{7})}{\Gamma(1+\frac{2l}{7})} (1+i\omegat S_l), \\
X_2&\equiv \frac{\Gamma(-1-\frac{2l}{7})}{\Gamma(-\frac{8}{7}-\frac{l}{7})\Gamma(\frac{1}{7}-\frac{l}{7})}\frac{\Gamma(\frac{6}{7}+\frac{l}{7})\Gamma(\frac{15}{7}+\frac{l}{7})}{\Gamma(3+\frac{2l}{7})} (1+i\omegat S_l), \\
S_l&\equiv \frac{\pi \sin(\frac{2l}{7}\pi)}{\sin(\frac{l-1}{7}\pi)\sin(\frac{l+1}{7}\pi)}.
\end{align}

The on-shell action for the vector modes is
\begin{align}
2\kappa^2S_{\mathrm{on-shell}}&=\Rt^{\frac{19}{5}}z_0^{-\frac{24}{5}}D_1\int_{u=u_c} dt \bigg[\frac{1}{2}z_0u^{-\frac{19}{14}}b^z\del_0 b^0+\bigg(\frac{2}{5}-\frac{7}{5}f^{-1}\bigg)u^{-\frac{12}{7}}(b^z)^2 \notag \\
&+\frac{7}{5}u^{-\frac{12}{7}}f\bigg((1+f)(b^0)^2-fub^0( b^0)'\bigg)\bigg] \notag \\
&-\frac{7}{5}\Rt z_0^{-2}D_1\int_{u=u_c} dt u^{-\frac{5}{7}}f b^0 \ahat +\frac{7}{5} \Rt^{-\frac{9}{5}}z_0^{\frac{4}{5}}D_1\int_{u=u_c}dt fu^{\frac{9}{7}}\ahat \ahat'. \label{eq:vectoronshell}
\end{align}
Derivation of the on-shell action is given in  Appendix \ref{sec:vectoronshell}.
The first line of (\ref{eq:vectoronshell}) is of the order of $\omegat^2$ because $b^z$ is proportional to $\omegat$ from (\ref{eq:bza1a2}). 
Since we only consider the solutions of the equations of motion to the linear order of $\omegat$, we neglect these terms.
Suppressing the contact terms of  (\ref{eq:vectoronshell}), the on-shell action which we should analyze is
\begin{align}
2\kappa^2S_{\mathrm{on-shell}}&=-\frac{7}{5}\Rt^{\frac{19}{5}}z_0^{-\frac{24}{5}}D_1\int_{u=u_c} dt f^2u^{-\frac{5}{7}}b^0( b^0)' +\frac{7}{5} \Rt^{-\frac{9}{5}}z_0^{\frac{4}{5}}D_1\int_{u=u_c}dt fu^{\frac{9}{7}}\ahat \ahat'. \label{eq:onsehllvector}
\end{align}

To evaluate the on-shell action, we need to express $(b^0)'$ and $\ahat'$ in terms of  $\bbar^0\equiv b^0(u_c)$ and $\abar\equiv a(u_c)$. In order to do this, we define the following functions:
\begin{align}
G_1&=u^{\frac{8}{7}+\frac{l}{7}}(u^{-\frac{1}{7}-\frac{l}{7}}F_1)', \\
\Gtilde_1&=u^{\frac{8}{7}-\frac{l}{7}}(u^{-\frac{1}{7}+\frac{l}{7}}\Ftilde_1)', \\
G_2&=u^{\frac{15}{7}+\frac{l}{7}}(u^{-\frac{8}{7}-\frac{l}{7}}F_2)', \\
\Gtilde_2&=u^{\frac{1}{7}-\frac{l}{7}}(u^{\frac{6}{7}+\frac{l}{7}}\Ftilde_2)', \\
H_1&=u^{\frac{15}{7}+\frac{l}{7}}(u^{-\frac{8}{7}-\frac{l}{7}}G_1)', \\
\Htilde_1&=u^{\frac{15}{7}-\frac{l}{7}}(u^{-\frac{8}{7}+\frac{l}{7}}\Gtilde_1)', \\
H_2&=u^{\frac{22}{7}+\frac{l}{7}}(u^{-\frac{15}{7}-\frac{l}{7}}G_2)', \\
\Htilde_2&=u^{\frac{8}{7}-\frac{l}{7}}(u^{-\frac{1}{7}+\frac{l}{7}}\Gtilde_2)', 
\end{align}
and
\begin{align}
\Fscr_1&=u_c^{-\frac{1}{7}-\frac{l}{7}}(F_1(u_c)-X_1u_c^{\frac{2l}{7}}\Ftilde_1(u_c)), \\
\Fscr_2&=u_c^{-\frac{8}{7}-\frac{l}{7}}(F_2(u_c)-X_2u_c^{2+\frac{2l}{7}}\Ftilde_2(u_c)), \\
\Gscr_1&=u_c^{-\frac{8}{7}-\frac{l}{7}}(G_1(u_c)-X_1u_c^{\frac{2l}{7}}\Gtilde_1(u_c)), \\
\Gscr_2&=u_c^{-\frac{15}{7}-\frac{l}{7}}(G_2(u_c)-X_2u_c^{2+\frac{2l}{7}}\Gtilde_2(u_c)), \\
\Hscr_1&=u_c^{-\frac{15}{7}-\frac{l}{7}}(H_1(u_c)-X_1u_c^{\frac{2l}{7}}\Htilde_1(u_c)), \\
\Hscr_2&=u_c^{-\frac{22}{7}-\frac{l}{7}}(H_2(u_c)-X_2u_c^{2+\frac{2l}{7}}\Htilde_2(u_c)).
\end{align}
Then, one finds
\begin{align}
\ahat_{1,2}'(u_c)&=\abar_{1,2}\frac{\Gscr_{1,2}}{\Fscr_{1,2}}, \label{eq:delahat1} \\
\ahat_{1,2}''(u_c)&=\abar_{1,2}\frac{\Hscr_{1,2}}{\Fscr_{1,2}}.
\end{align}
From (\ref{eq:ahat12}) and (\ref{eq:b012}), $\abar_1$ and $\abar_2$ are
\begin{align}
\abar_1&=\frac{\Fscr_1}{Q}\bigg(-\frac{\Gscr_2}{l+8}g_s\abar -\frac{1}{7}\bigg(\frac{\Rt}{z_0}\bigg)^{\frac{14}{5}}u_c^{-2}\Fscr_2 \bbar^0\bigg), \label{eq:abar1tobbar} \\
\abar_2&=\frac{\Fscr_2}{Q}\bigg(-\frac{\Gscr_1}{l-1}g_s\abar+\frac{1}{7}\bigg(\frac{\Rt}{z_0}\bigg)^{\frac{14}{5}}u_c^{-2}\Fscr_1 \bbar^0\bigg), \label{eq:abar2tobbar}
\end{align}
where
\begin{align}
Q&=\frac{\Fscr_1\Gscr_2}{l+8}+\frac{\Fscr_2\Gscr_1}{l-1}. \label{eq:Q}
\end{align}
Taking the derivative of (\ref{eq:ahat12}) and (\ref{eq:b012}) with respect to $u$ and using (\ref{eq:delahat1})--(\ref{eq:abar2tobbar}), we find
\begin{align}
(b^0)'|_{u=u_c}&=\frac{2}{u_c}\bbar^0+\frac{\bbar^0}{Q}\bigg(\frac{\Fscr_2\Hscr_1}{l-1}+\frac{\Fscr_1\Hscr_2}{l+8}\bigg)-\frac{7u_c^2 g_s\abar}{(l-1)(l+8)Q}\bigg(\frac{z_0}{\Rt}\bigg)^{\frac{14}{5}}(\Gscr_1\Hscr_2-\Hscr_1\Gscr_2), \label{eq:b0prime}  \\
\ahat'|_{u=u_c}&=\frac{\Fscr_1\Gscr_2-\Fscr_2\Gscr_1}{7u_c^2 Q}\bigg(\frac{\Rt}{z_0}\bigg)^{\frac{14}{5}}\bbar^0-\frac{2l+7}{(l+8)(l-1)}\frac{\Gscr_1\Gscr_2}{Q}g_s\abar. \label{eq:aprime}
\end{align}
Since the first term of (\ref{eq:b0prime}) becomes a contact term in the on-shell action, we will suppress it.
In terms of $\Fscr_{1,2}, \Gscr_{1,2}, \Hscr_{1,2}$, and $Q$, 
the on-shell action (\ref{eq:onsehllvector}) is expressed as 
\begin{align}
2\kappa^2 S_{\mathrm{on-shell}}&=-\frac{7}{5}\Rt^{\frac{19}{5}}z_0^{-\frac{24}{5}}D_1\int_{u=u_c} \frac{d\omega}{2\pi} u^{-\frac{5}{7}}(1-u)^2\frac{1}{Q}\bigg(\frac{\Fscr_2\Hscr_1}{l-1}+\frac{\Fscr_1\Hscr_2}{l+8}\bigg)\bbar^0(-\omega)\bbar^0(\omega) \notag \\
&+\frac{49}{5}g_s\Rt z_0^{-2}D_1\int_{u=u_c}\frac{d\omega}{2\pi} u^{\frac{9}{7}}(1-u)^2\frac{\Gscr_1\Hscr_2-\Hscr_1\Gscr_2}{(l-1)(l+8)Q}\bbar^0(-\omega)\abar(\omega) \notag \\
&-\frac{1}{5}g_s\Rt z_0^{-2}D_1\int_{u=u_c}\frac{d\omega}{2\pi}  u^{-\frac{5}{7}}(1-u)\frac{\Fscr_1\Gscr_2-\Fscr_2\Gscr_1}{Q}\bbar^0(-\omega)\abar(\omega) \notag \\
&+\frac{7}{5}g_s^2\Rt^{-\frac{9}{5}}z_0^{\frac{4}{5}}D_1\int_{u=u_c} \frac{d\omega}{2\pi} u^{\frac{9}{7}}(1-u)\frac{2l+7}{(l-1)(l+8)}\frac{\Gscr_1\Gscr_2}{Q}\abar(-\omega)\abar(\omega). \label{eq:generalvectoronshell}
\end{align}

\subsection{The case of $u_c\simeq 1$} \label{sec:uc1}
In the previous section, we have calculated the on-shell action 
on the cutoff surface at $u_c$. 
Since the solutions are expressed in terms of the hypergeometric functions, 
it is difficult to discuss properties of the linear response for arbitrary $u_c$. 
Hence, we focus on two regions, $u_c\simeq 1$ and $u_c\simeq 0$. 
We first consider the case of $u_c\simeq 1$ in which the cutoff surface is near the horizon. 
This corresponds to putting a cutoff at the low energy scale in the Matrix theory side. 
Following  \cite{Bredberg:2010ky, Matsuo:2012pi}, we evaluate 
the linear responses of Matrix theory in terms of the proper quantities on the cutoff surface. According to \cite{Bredberg:2010ky, Matsuo:2012pi}, the stress tensor and R-R 1-form current are given by 
\begin{align}
\Tcal^{\mu\nu}&=\frac{\delta S_{\mathrm{on-shell}}}{\sqrt{-\gamma}\delta \gamma_{\mu\nu}}, \label{eq:stressbrown} \\
\Jcal^{\mu}&=\frac{\delta S_{\mathrm{on-shell}}}{\sqrt{-\gamma}\delta \bar{A}_{\mu}}, \label{eq:currentbrown}
\end{align}
where $\gamma_{\mu\nu}$ and $\bar{A}_{\mu}$  are the induced metric and R-R 1-form on the cutoff surface, respectively. 
These expressions can be understood in terms of the quasilocal charges. 
In fact, the expression (\ref{eq:stressbrown}) is the same as the definition of the Brown-York stress tensor \cite{Brown:1992br}.

\subsubsection{Tensor mode}
Here, we consider the tensor mode. 
Expanding (\ref{eq:tensoronshellgen}) around $u_c=1$, the on-shell action becomes
\begin{align}
S_{\mathrm{on-shell}}&=\frac{1}{16\pi G}\frac{1}{2}\bigg(\frac{\Rt}{z_0}\bigg)^{\frac{9}{5}} D_2\int_{u_c\simeq 1}\frac{d\omega}{2\pi}i\omega  \bbar(-\omega)\bbar(\omega),
\end{align}
up to the linear order of $\omega$. Here, we have suppressed the contact terms in the action.  

Let us define the proper frequency as
\begin{align}
\frakw=\frac{\omega}{\sqrt{-g_{00}}}.
\end{align}
Then, the on-shell action can be written as
\begin{align}
S_{\mathrm{on-shell}}&=\frac{1}{16\pi G}\frac{1}{2}\bigg(\frac{\Rt}{z_0}\bigg)^{\frac{9}{5}} D_2\int_{u_c\simeq 1}\frac{d\omega}{2\pi}\sqrt{-g_{00}}i\frakw \bbar(-\omega)\bbar(\omega).
\end{align}
The tensor mode of the metric perturbation $\bbar(\omega)$ is coupled to the tensor mode of the stress tensor $\Tcal(\omega)$  in Matrix theory.\footnote{The tensor mode of the stress tensor is the Fourier coefficient of the tensor harmonics $Y^{ij}$ in the spherical harmonic expansion of the stress tensor $\Tcal^{ij}$.} According to  section \ref{sec:nearextcorr} and (\ref{eq:stressbrown}), the linear response of the stress tensor is
\begin{align}
\Tcal(\omega)&=\frac{1}{16\pi G}\bigg(\frac{r_0}{R}\bigg)^{\frac{9}{2}}i\frakw \bbar(\omega), \label{eq:tensortu1}
\end{align}
which is the same as  (\ref{eq:hydrotensor}) with
\begin{align}
\eta&=\frac{1}{16\pi G}\bigg(\frac{r_0}{R}\bigg)^{\frac{9}{2}}. \label{eq:shearviscosity}
\end{align}
Therefore, the linear response obeys the hydrodynamics on $S^8$ when $u_c\simeq 1$. Since the entropy density on the horizon in AdS frame is
\begin{align}
s=\frac{S_{BH}}{V_8}=\frac{1}{4G}\bigg(\frac{r_0}{R}\bigg)^{\frac{9}{2}},
\end{align}
we find\footnote{The dimensionless transport coefficients such as $\eta/s$ do not depend on a choice of the frame.}
\begin{align}
\frac{\eta}{s}=\frac{1}{4\pi}, \label{eq:etaovers}
\end{align}
which is the  same  as the shear viscosity to entropy density ratio in the membrane paradigm \cite{Thorne:1986iy} or the AdS/CFT correspondence \cite{Policastro:2002se}.




\subsubsection{Vector modes}
Next, we consider the vector mode for $u_c\simeq 1$. 
Expanding  $\Fscr_{1,2}, \Gscr_{1,2}, \Hscr_{1,2}$, and $Q$ around $u_c=1$ and suppressing the contact terms, the on-shell action  becomes%
\footnote{Strictly speaking, we should neglect the  $\Ocal(\frakw^2)$ term in the numerator of the last term in (\ref{eq:u1vectoronshell}), since we have calculated only to linear order of $\frakw$. 
Although this term could possibly receive corrections if we calculate $\Ocal(\frakw^2)$ contributions, 
we can see that this term is consistent with charged fluid.}
\begin{align}
2\kappa^2 S_{\mathrm{on-shell}}&=\frac{1}{2}\bigg(\frac{\Rt}{z_0}\bigg)^{\frac{9}{5}} D_1\int_{u_c\simeq 1}\frac{d\omega}{2\pi} \sqrt{-g_{00}}\frac{\frac{(l+8)(l-1)}{R^2}}{i\frakw-\Dcal\big(\frac{(l+8)(l-1)}{R^2}-\frac{14(2l^2+14l-7)}{(2l+7)S_lR^2}\big)}\bbar^{\tilde{0}}(-\omega)\bbar_{\tilde{0}}(\omega), \notag \\
&-\frac{7g_s}{R}D_1 \int_{u_c \simeq 1}\frac{d\omega}{2\pi} \sqrt{-g_{00}} \frac{i\frakw +\Dcal\frac{14(l^2+7l+1)}{(2l+7)S_lR^2}}{i\frakw-\Dcal\big(\frac{(l+8)(l-1)}{R^2}-\frac{14(2l^2+14l-7)}{(2l+7)S_lR^2}\big)}\bbar_{\tilde{0}}(-\omega)\abar(\omega) \notag \\
&+\frac{g_s^2}{2}\bigg(\frac{z_0}{\Rt}\bigg)^{\frac{9}{5}}D_1\int_{u_c\simeq 1} \frac{d\omega}{2\pi} \sqrt{-g_{00}}\abar(-\omega)\abar(\omega) \notag \\
&\cdot \frac{\Dcal^2\big(\frac{(l-1)(l+1)}{R^2}-\frac{14l}{S_lR^2}\big)\big(\frac{(l+6)(l+8)}{R^2}-\frac{14(l+7)}{S_l R^2}\big)-i\frakw \Dcal\big(\frac{2l^2+14 l+47}{R^2}-\frac{14(2l+7)}{S_lR^2}\big)-\frakw^2}{i\frakw-\Dcal\big(\frac{(l+8)(l-1)}{R^2}-\frac{14(2l^2+14l-7)}{(2l+7)S_lR^2}\big)}, \label{eq:u1vectoronshell}
\end{align}
where  $\tilde{0}$ is the proper time index (namely, $\bbar_{\tilde{0}}=\bbar_0/\sqrt{-g_{00}}$) and
\begin{align}
\Dcal&\equiv \frac{1}{4\pi T}=\frac{\sqrt{-g_{00}}}{4\pi T_H}.
\end{align}
Here, $T$ is the proper temperature. 

The vector modes of the source fields $\bbar^0(\omega)$ and $\abar(\omega)$ are coupled to the vector modes of the stress tensor and R-R 1-form current in Matrix theory, respectively.
According to section \ref{sec:nearextcorr} and (\ref{eq:stressbrown})-(\ref{eq:currentbrown}), the linear responses of the vector modes of the stress tensor and  R-R 1-form current are
\begin{align}
\Tcal^{\tilde{0}}&=\frac{1}{16\pi G}\bigg(\frac{r_0}{R}\bigg)^{\frac{9}{2}} \frac{\frac{(l+8)(l-1)}{R^2}}{i\frakw-\Dcal\big(\frac{(l+8)(l-1)}{R^2}-\frac{14(2l^2+14l-7)}{(2l+7)S_lR^2}\big)}\bbar^{\tilde{0}} \notag \\
&-\frac{7g_s}{16\pi G R}  \frac{i\frakw +\Dcal\frac{14(l^2+7l+1)}{(2l+7)S_lR^2}}{i\frakw-\Dcal\big(\frac{(l+8)(l-1)}{R^2}-\frac{14(2l^2+14l-7)}{(2l+7)S_lR^2}\big)}\abar, \label{eq:vectorstressu1} \\
\Jcal&=\frac{g_s^2}{16\pi G}\bigg(\frac{R}{r_0}\bigg)^{\frac{9}{2}}i\frakw \abar-\frac{49 \Dcal g_s^2}{16\pi G R^2}\bigg(\frac{R}{r_0}\bigg)^{\frac{9}{2}}\frac{ i \frakw-V}{i\frakw-\Dcal\big(\frac{(l+8)(l-1)}{R^2}-\frac{14(2l^2+14l-7)}{(2l+7)S_lR^2}\big)}\abar \notag \\
&+\frac{7g_s}{16\pi G R}  \frac{i\frakw +\Dcal\frac{14(l^2+7l+1)}{(2l+7)S_lR^2}}{i\frakw-\Dcal\big(\frac{(l+8)(l-1)}{R^2}-\frac{14(2l^2+14l-7)}{(2l+7)S_lR^2}\big)}\bbar^{\tilde{0}}, \label{eq:vectorcurrentu1}
\end{align}
where we have suppressed the contact terms and
\begin{align}
V=\frac{18}{(2l+7)S_l}\bigg(i\frakw -\Dcal\frac{l^2+7l-1}{R^2}\bigg)+\frac{28 \Dcal}{(2l+7)^2S_l^2}\frac{11l^2+77l-7}{R^2}.
\end{align}

Comparing the linear responses to the hydrodynamic stress tensor and current on $S^8$ with radius $R$, which are given by  (\ref{eq:hydrot0}) and (\ref{eq:hydroj}), we find the following:
\begin{itemize}
\item Compared to the diffusion pole in  (\ref{eq:hydrot0}) and (\ref{eq:hydroj}), 
the denominators in (\ref{eq:vectorstressu1}) and (\ref{eq:vectorcurrentu1}) possess 
an extra term, $\frac{14(2l^2+14l-7)}{(2l+7)S_lR^2}$. 
However, the quantity $\Dcal$ matches with the diffusion constant (\ref{eq:diffusionconst}). In fact, using the thermodynamic relation, $\ebar+\pbar=Ts$, and $\eta/s=1/4\pi$, the diffusion constant becomes
\begin{align}
D=\frac{\eta}{\ebar+\pbar}=\frac{\eta}{s}\cdot \frac{1}{T}=\frac{1}{4\pi T}=\Dcal.
\end{align}
\item Except for the extra term in the denominator, the first term of (\ref{eq:vectorstressu1}) agrees with the first term of  (\ref{eq:hydrot0}) because the shear viscosity is given by (\ref{eq:shearviscosity}).
\item 
Except for the extra terms in the denominator and numerator, the second term of (\ref{eq:vectorstressu1}) and the third term of  (\ref{eq:vectorcurrentu1}) agree with the second term of (\ref{eq:hydrot0}) and the third term of (\ref{eq:hydroj}), respectively, because from (\ref{eq:totalcharge}), the charge density on the horizon in AdS frame is
\begin{align}
\nbar=\frac{q}{V_8}=\frac{7g_s}{16\pi GR}.
\end{align}
\item
Except for the extra terms in the denominator and numerator, the second term of  (\ref{eq:vectorcurrentu1})  agrees with the second term of (\ref{eq:hydroj}) because 
\begin{align}
\frac{\nbar^2}{\ebar+\pbar}=\bigg(\frac{7g_s}{16 \pi G R}\bigg)^2\frac{1}{Ts}=\frac{49\Dcal g_s^2}{16\pi GR^2}\bigg(\frac{R}{r_0}\bigg)^{\frac{9}{2}}.
\end{align}
\item
The first term of (\ref{eq:vectorcurrentu1}) agrees with the first term of (\ref{eq:hydroj}) if
\begin{align}
\sigma=\frac{g_s^2}{16\pi G}\bigg(\frac{R}{r_0}\bigg)^{\frac{9}{2}}.
\end{align}
\item
The extra terms which appear in (\ref{eq:vectorstressu1}) and (\ref{eq:vectorcurrentu1}) are decoupled if we take $l$ as large with $l/R$ fixed ($S_l^{-1}$ is of the order of $1$). This means that  the linear responses of the vector modes locally obey the hydrodynamics.
\end{itemize}

Although we have obtained charged fluid, the fluid should have
universal structure near the horizon.
Such universal structures appear if we take the Rindler limit.
In the next subsection, we see that by taking a Rindler limit, (\ref{eq:vectorstressu1}) and (\ref{eq:vectorcurrentu1}) become the hydrodynamic stress tensor and current with $\nbar=0$ in the 8-dimensional flat space.

\subsubsection{Rindler limit}
Let us look at a local region of $S^8$, which can be approximated by $\mathbf{R}^8$. 
Then, the metric of the 8-dimensional space is replaced by
\begin{align}
R^2d\Omega_8^2 \to dx^idx_i.
\end{align}
The magnitude of the momentum in the flat 8-dimensional space is 
\begin{align}
k=\frac{l}{R}.
\end{align}

The Rindler limit is defined as follows:
setting
\begin{align}
1-u&=\bigg(\frac{7}{5\Rt}\bigg)^2\varepsilon^2\rhat^2, \\
x^i&=\varepsilon \hat{x}^i, \label{eq:rindlerxi}
\end{align}
and sending $\varepsilon\to 0$, the metric (\ref{eq:nearext1}) becomes conformal to the Rindler metric,
\begin{align}
ds_s^2&=\varepsilon^2d\hat{s}^2, \\
d\hat{s}^2&=-\kappa^2\rhat^2dt^2+d\rhat^2+d\xhat^id\xhat_i,
\end{align}
where $\kappa=2\pi T_H$ gives the Hawking temperature of the Rindler spacetime, $T_H$. 
Since we have magnified a small region in $S^8$, 
the proper frequency, momentum, and proper temperature are also rescaled. 
Those in the Rindler spacetime are related to the original ones as
\begin{align}
\hat{\frakw}&=\varepsilon \frakw, \\
\khat^i&=\varepsilon k^i, \\
\That&=\varepsilon T.
\end{align}
The $(0,i)$ component of the metric perturbation  in the Rindler spacetime  is related to the original one as
\begin{align}
h_{0i}=\varepsilon \hat{h}_{0i},
\end{align}
because $t$ is not rescaled by $\varepsilon$ \cite{Matsuo:2012pi}. The Newton constant $G\sim g_s^2l_s^8$ is also rescaled as $G=\varepsilon^8 \hat{G}$, where $\hat{G}$ is the Newton constant in the Rindler spacetime because the string length in the Rindler spacetime is $\hat{l}_s=l_s /\varepsilon$.
In this limit, the stress tensor (\ref{eq:vectorstressu1}) and 
the R-R 1-form current (\ref{eq:vectorcurrentu1}) become
\begin{align}
\Tcal^{\tilde{0}i}&=\frac{1}{16\pi \hat{G}\varepsilon^9}\bigg(\frac{r_0}{R}\bigg)^{\frac{9}{2}}\frac{\khat^2}{i\hat{\frakw}-\hat{\Dcal} \khat^2}\hat{h}^{\tilde{0}i}-\frac{7g_s}{16\pi \hat{G} R\varepsilon^8} \frac{i\hat{\frakw}}{i\hat{\frakw}-\hat{\Dcal} \hat{k}^2}\delta A^i, \label{eq:vectorstressu2} \\
\Jcal^i&=\frac{g_s^2}{16\pi \hat{G}\varepsilon^9}\bigg(\frac{R}{r_0}\bigg)^{\frac{9}{2}}
 i\hat{\frakw} \delta A^i-\frac{49\hat{\Dcal}g_s^2}{16\pi \hat{G}R^2 \varepsilon^7}\bigg(\frac{R}{r_0}\bigg)^{\frac{9}{2}}\frac{i\hat{\frakw}}{i\hat{\frakw}-\hat{\Dcal}\hat{k}^2} \delta A^i +\frac{7g_s}{16\pi \hat{G} R\varepsilon^8}  \frac{i\hat{\frakw} }{i\hat{\frakw}-\hat{\Dcal} \hat{k}^2}\hat{h}^{\tilde{0}i}, \label{eq:vectorcurrentu2}
\end{align}
where $\hat{h}^{\tilde{0}i}=\kappa \hat{r}\hat{h}^{0i}$ and  $\hat{\Dcal}\equiv \frac{1}{4\pi \hat{T}}$ . We have omitted the bar ($\bar{~}$) which denotes  the perturbations on the cutoff surface. Since the stress tensor and current in the Rindler spacetime are related to the original ones as
\begin{align}
\hat{\Tcal}^{\tilde{0}i}&=\varepsilon^9 \Tcal^{\tilde{0}i}, \\
\hat{\Jcal}^{i}&=\varepsilon^9 \Jcal^{i},
\end{align}
we find
\begin{align}
\hat{\Tcal}^{\tilde{0}i}&=\frac{1}{16\pi \hat{G}}\bigg(\frac{r_0}{R}\bigg)^{\frac{9}{2}}\frac{\khat^2}{i\hat{\frakw}-\hat{\Dcal} \khat^2}\hat{h}^{\tilde{0}i}, \label{eq:rindlerlimitstress} \\
\hat{\Jcal}^i&=\frac{g_s^2}{16\pi \hat{G}}\bigg(\frac{R}{r_0}\bigg)^{\frac{9}{2}}
 i\hat{\frakw} \delta A^i. \label{eq:rindlerlimitcurrent}
\end{align}
Note that the second term in (\ref{eq:vectorstressu2}) and the second and third terms in (\ref{eq:vectorcurrentu2}) are decoupled in the limit of $\varepsilon\to 0$. Therefore, in the Rindler limit, (\ref{eq:vectorstressu1}) and (\ref{eq:vectorcurrentu1}) exactly match with the hydrodynamic stress tensor and current on $\mathbf{R}^8$ with no charge density. This result is consistent with the previous works on a Rindler fluid \cite{Bredberg:2010ky,Matsuo:2012pi}. In a Rindler fluid, there is no charge density because the Rindler metric is a solution of the vacuum Einstein equation.

\subsection{The case of $u_c\simeq 0$} \label{sec:uc0}
We consider the case in which the cutoff surface is far from the horizon. At first, we calculate the linear responses in terms of the proper quantities on the cutoff surface as in the previous section. 

When we do not put the cutoff surface but consider the asymptotic boundary at $r\to\infty$, 
or equivalently, in the limit of $u_c=0$, 
the divergent warp factor in the gravity side should be excluded from the correspondence  \cite{Gubser:1998bc,Witten:1998qj}. In our case, since the dual geometry of Matrix theory is essentially $AdS_2$ ($S^8$ is interpreted as the internal space in Matrix theory), we have to care about the time-time component of the metric.
We take into account  the metric of Matrix theory and obtain the linear responses of the stress tensor and R-R 1-form current. Then, we compare the linear responses with the hydrodynamic stress tensor and current on $S^8$ and discuss the differences between them.
\subsubsection{Tensor mode}
Here, we consider the tensor mode. 
Expanding (\ref{eq:tensoronshellgen}) around $u_c=0$,  the on-shell action becomes
\begin{align}
S_{\mathrm{on-shell}}&=\frac{1}{16\pi G}\frac{1}{2}\bigg(\frac{r_0}{R}\bigg)^{\frac{9}{2}}D_2\int\frac{d\omega}{2\pi} i\omega \frac{\Gamma(1+\frac{l}{7})^4}{\Gamma(1+\frac{2l}{7})^2} u_c^{\frac{2}{7}l}\bbar(-\omega)\bbar(\omega),\label{eq:finaltensoru0}
\end{align}
to the linear order of $\omega$.
Expressing this in terms of the proper quantities, we find
\begin{align}
S_{\mathrm{on-shell}}&=\frac{1}{16\pi G}\frac{1}{2}\bigg(\frac{r_0}{R}\bigg)^{\frac{9}{2}}D_2\int\frac{d\omega}{2\pi} \sqrt{-g_{00}}i\frakw \frac{\Gamma(1+\frac{l}{7})^4}{\Gamma(1+\frac{2l}{7})^2} u_c^{\frac{2}{7}l}\bbar(-\omega)\bbar(\omega). \label{eq:finaltensoru1}
\end{align}
Although the factor $u_c^{\frac{2l}{7}}$ is usually absorbed into the field redefinition $\bbar\to u_c^{-\frac{l}{7}}\bbar$, we do not consider such a  wave function renormalization because it does not change our conclusion. Namely, $\bbar$ is a bare field in an energy scale  which is determined by $u_c$.

Therefore, the linear response of the tensor mode of the stress tensor in terms of the proper quantities   is
\begin{align}
\Tcal(\omega)=\frac{1}{16\pi G}\bigg(\frac{r_0}{R}\bigg)^{\frac{9}{2}}i\frakw \frac{\Gamma(1+\frac{l}{7})^4}{\Gamma(1+\frac{2l}{7})^2} u_c^{\frac{2}{7}l}\bbar(\omega).
\end{align}
Comparing this with (\ref{eq:tensortu1}), we find the extra factor $\Gamma(1+\frac{l}{7})^4/\Gamma(1+\frac{2l}{7})^2$ except for the factor $u_c^{\frac{2l}{7}}$, which could be absorbed into the field redefinition of $\bbar$.

Taking into account  the metric of Matrix theory, the linear response becomes
\begin{align}
\Tcal(\omega)=\frac{1}{16\pi G}\bigg(\frac{r_0}{R}\bigg)^{\frac{9}{2}}i\omega  \frac{\Gamma(1+\frac{l}{7})^4}{\Gamma(1+\frac{2l}{7})^2} u_c^{\frac{2}{7}l}\bbar(\omega). \label{eq:tcalfinalu0}
\end{align}
This is  different from the hydrodynamic stress tensor (\ref{eq:hydrotensor})  even if we absorb the factor $u_c^{\frac{2l}{7}}$ into the field redefinition because the shear viscosity $\eta$ does not depend on $l$. Therefore, for the tensor mode, the linear response of the D0-branes cannot be explained by the hydrodynamics when the cutoff surface is far from the horizon.

\subsubsection{Vector modes}
Here, we consider the vector modes. 
Expanding (\ref{eq:generalvectoronshell}) around $u_c=0$, the on-shell action becomes
\begin{align}
2\kappa^2 S_{\mathrm{on-shell}}&=-\frac{7}{5}\Rt^{\frac{19}{5}}z_0^{-\frac{24}{5}}D_1\int \frac{d\omega}{2\pi} \bbar^0(-\omega)\bbar^0(\omega)u_c^{-\frac{19}{7}-\frac{2l}{7}}\frac{2l+7}{18l(l-1)B^2} \notag \\
&\cdot \frac{2l^2+23l-7+\frac{2}{7}l(2l+7)(l-1)B^2u_c^{\frac{2l}{7}}i\omegat}{i\omegat -\frac{7(2l+7)}{9(l-1)B^2}u_c^{-1-\frac{2l}{7}}}, \notag \\
&-\frac{7}{5}g_s \Rt z_0^{-2} D_1\int \frac{d\omega}{2\pi} \bbar^0(-\omega)\abar(\omega)u_c^{-\frac{12}{7}-\frac{2l}{7}}\frac{2l+7}{9l(l-1)B^2} \notag \\
&\cdot \frac{7-\frac{2}{7}l(2l+7)(l-1)B^2u_c^{\frac{2l}{7}}i\omegat}{i\omegat -\frac{7(2l+7)}{9(l-1)B^2}u_c^{-1-\frac{2l}{7}}}, \notag \\
&+\frac{7}{5}g_s^2\Rt^{-\frac{9}{5}}z_0^{\frac{4}{5}}D_1\int\frac{d\omega}{2\pi}\abar(-\omega)\abar(\omega)u_c^{-\frac{5}{7}-\frac{2l}{7}}\frac{(2l+7)^2}{18l(l-1)B^2} \notag \\
&\cdot \frac{l+1-\frac{2}{7}l(l-1)B^2u_c^{\frac{2l}{7}}i\omegat}{i\omegat -\frac{7(2l+7)}{9(l-1)B^2}u_c^{-1-\frac{2l}{7}}},
\end{align}
to the linear order of $\omegat$, where
\begin{align}
B\equiv \frac{\Gamma(-\frac{1}{7}+\frac{l}{7})\Gamma(\frac{8}{7}+\frac{l}{7})}{\Gamma(1+\frac{2l}{7})}.
\end{align} 
According to  section \ref{sec:nearextcorr} and (\ref{eq:stressbrown})--(\ref{eq:currentbrown}), the linear responses of the vector modes of the stress tensor and R-R 1-form current in terms of the proper quantities are
\begin{align}
\Tcal^{\tilde{0}}&=\frac{1}{16\pi G}\bigg(\frac{r_0}{R}\bigg)^{\frac{9}{2}}u_c^{-\frac{9}{7}-\frac{2l}{7}}\frac{2l+7}{18l(l-1)B^2} \notag \\
&\cdot \frac{\frac{49}{R^2}(2l^2+23l-7)+\frac{2}{R}l(2l+7)(l-1)B^2i\frakw  u_c^{-\frac{5}{14}+\frac{2l}{7}}}{i\frakw -\frac{49 (2l+7)}{9(l-1)B^2R}u_c^{-\frac{9}{14}-\frac{2l}{7}}}\bbar^{\tilde{0}} \notag \\
&-\frac{7g_s}{16\pi GR}\frac{(2l+7)^2}{63u_c}\frac{i\frakw -\frac{343}{2l(2l+7)(l-1)B^2R}u_c^{\frac{5}{14}-\frac{2l}{7}}}{i\frakw -\frac{49 (2l+7)}{9(l-1)B^2R}u_c^{-\frac{9}{14}-\frac{2l}{7}}}\abar, \label{eq:stressvecfinal}  \\
\Jcal&=\frac{g_s^2}{16\pi G}\bigg(\frac{R}{r_0}\bigg)^{\frac{9}{2}}\frac{(2l+7)^2}{9}u_c^{-\frac{5}{14}}\frac{\frac{49(l+1)}{2l(l-1)B^2R^2}u_c^{\frac{5}{14}-\frac{2l}{7}}-\frac{i\frakw}{R}}{i\frakw -\frac{49 (2l+7)}{9(l-1)B^2R}u_c^{-\frac{9}{14}-\frac{2l}{7}}}\abar \notag \\
&+\frac{7g_s}{16\pi GR}\frac{(2l+7)^2}{63u_c}\frac{i\frakw -\frac{343}{2l(2l+7)(l-1)B^2R}u_c^{\frac{5}{14}-\frac{2l}{7}}}{i\frakw -\frac{49 (2l+7)}{9(l-1)B^2R}u_c^{-\frac{9}{14}-\frac{2l}{7}}}\bbar^{\tilde{0}}. \label{eq:currentvecfinal}
\end{align}
Taking into account  the metric of Matrix theory, the linear responses become
\begin{align}
\Tcal^0&=\frac{1}{16\pi G}\bigg(\frac{r_0}{R}\bigg)^{\frac{9}{2}}u_c^{-\frac{9}{7}-\frac{2l}{7}}\frac{2l+7}{18l(l-1)B^2} \notag \\
&\cdot \frac{\frac{49}{R^2}(2l^2+23l-7)+\frac{5z_0}{R^2}l(2l+7)(l-1)B^2u_c^{\frac{2l}{7}}i\omega}{i\omega -\frac{98(2l+7)}{45z_0(l-1)B^2}u_c^{-1-\frac{2l}{7}}}\bbar^0 \notag \\
&-\frac{7g_s}{16\pi GR}\frac{(2l+7)^2}{63u_c}\frac{i\omega -\frac{343}{5z_0l(2l+7)(l-1)B^2}u_c^{-\frac{2l}{7}}}{i\omega-\frac{98(2l+7)}{45z_0(l-1)B^2}u_c^{-1-\frac{2l}{7}}}\abar, \label{eq:stressvecfinale} \\
\Jcal&=\frac{g_s^2}{16\pi G}\bigg(\frac{R}{r_0}\bigg)^{\frac{9}{2}}u_c^{-\frac{5}{7}}\frac{2(2l+7)^2}{45}\frac{\frac{49(l+1)}{5l(l-1)z_0^2 B^2 }u_c^{-\frac{2l}{7}}-\frac{i\omega}{z_0}}{i\omega-\frac{98(2l+7)}{45z_0(l-1)B^2}u_c^{-1-\frac{2l}{7}}}\abar \notag \\
&+\frac{7g_s}{16\pi GR}\frac{(2l+7)^2}{63u_c}\frac{i\omega -\frac{343}{5z_0l(2l+7)(l-1)B^2}u_c^{-\frac{2l}{7}}}{i\omega-\frac{98(2l+7)}{45z_0(l-1)B^2}u_c^{-1-\frac{2l}{7}}}\bbar^0. \label{eq:currentvecfinale}
\end{align}
These expressions are quite different from the linear responses in the case of $u_c\simeq 1$ or the hydrodynamic stress tensor and current on $S^8$. Especially, there is no  pole structure in (\ref{eq:stressvecfinale}) and (\ref{eq:currentvecfinale}) (or (\ref{eq:stressvecfinal}) and (\ref{eq:currentvecfinal})) in  the low frequency region because in any $l \geq 1$, the factor $u_c^{-1-\frac{2l}{7}}$ (or $u_c^{-\frac{9}{14}-\frac{2l}{7}}$) is very large when $u_c \simeq 0$. It is important to note that this fact is independent of the field redefinitions of $\bbar^0$ and $\abar$. On the other hand, in hydrodynamics, there is a diffusion pole  in the low frequency region.

\begin{figure}
\begin{center}
\includegraphics[scale=1.2]{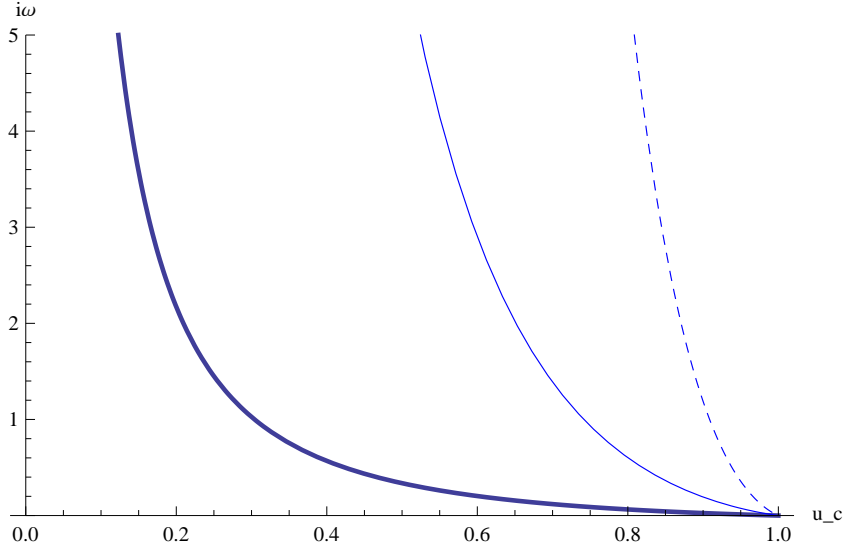}
\end{center}
\caption{The right-hand side of (\ref{eq:genucpole}) is plotted against $u_c$ when $l=2$ (bold line), $l=5$ (normal line), and $l=10$ (dashed line). }
\label{fig:genucpole}
\end{figure}

In order to look at the pole structure in the vector modes, 
let us consider the denominators in the above expressions for arbitrary $u_c$.  
In general $u_c$, the denominators of the linear responses in the vector modes vanish when (\ref{eq:Q}) equals  zero. Namely, it is when
\begin{align}
i\omegat&=\frac{K_1-B_1^{-1}\Btilde_1K_2-B_2^{-1}\Btilde_2K_3+B_1^{-1}\Btilde_1B_2^{-1}\Btilde_2K_4}{S_l(B_1^{-1}\Btilde_1K_2+B_2^{-1}\Btilde_2 K_3-2B_1^{-1}\Btilde_1B_2^{-1}\Btilde_2K_4)}, \label{eq:genucpole}
\end{align}
where
\begin{align}
K_1&=\frac{F_1G_2}{l+8}+\frac{F_2G_1}{l-1}, \\
K_2&=u^{\frac{2l}{7}}\bigg(\frac{\Ftilde_1G_2}{l+8}+\frac{F_2\Gtilde_1}{l-1}\bigg), \\
K_3&=u^{2+\frac{2l}{7}}\bigg(\frac{F_1\Gtilde_2}{l+8}+\frac{\Ftilde_2G_1}{l-1}\bigg), \\
K_4&=u^{2+\frac{4l}{7}}\bigg(\frac{\Ftilde_1\Gtilde_2}{l+8}+\frac{\Ftilde_2\Gtilde_1}{l-1}\bigg),
\end{align}
and 
\begin{align}
B_1&=\frac{\Gamma(-\frac{l+1}{7})\Gamma(\frac{8-l}{7})}{\Gamma(1-\frac{2l}{7})}, \\
\Btilde_1&=\frac{\Gamma(\frac{l-1}{7})\Gamma(\frac{l+8}{7})}{\Gamma(1+\frac{2l}{7})}, \\
B_2&=\frac{\Gamma(-\frac{l+8}{7})\Gamma(\frac{1-l}{7})}{\Gamma(-1-\frac{2l}{7})}, \\
\Btilde_2&=\frac{\Gamma(\frac{l+6}{7})\Gamma(\frac{l+15}{7})}{\Gamma(3+\frac{2l}{7})}.
\end{align}
Figure \ref{fig:genucpole} shows the value of the right-hand side of (\ref{eq:genucpole}) against $u_c$. Since $\omegat\simeq \omega/T_H=\frakw/T$,  the left-hand side of (\ref{eq:genucpole}) does not depend on the redshift. When $u_c\simeq 1$, the value of the right-hand side of (\ref{eq:genucpole})  is close to zero for any $l$. Therefore, we can find a  pole  structure in the  low frequency region when the cutoff surface is close to the horizon. However, as $u_c$ approaches  zero, the value of the right-hand side of (\ref{eq:genucpole}) becomes large and  exceeds 1. Therefore, within the low frequency approximation, we cannot find the pole structure when the cutoff surface is far from the horizon. 

\section{Summary and comments}
We have studied the linear responses of the near-extremal D0-branes in the low frequency region by using the gauge/gravity correspondence.  We have analyzed the tensor mode and vector modes. We have found that when the cutoff surface, on which Matrix theory is defined, is close to the horizon, the linear responses of the stress tensor and R-R 1-form current in Matrix theory take forms similar to the hydrodynamic stress tensor and current on $S^8$ with radius $R$. 
By taking the Rindler limit, the linear responses of Matrix theory exactly agree with the hydrodynamic stress tensor and current on $\mathbf{R}^8$, which is consistent with the previous result on a Rindler fluid \cite{Bredberg:2010ky, Matsuo:2012pi}. This is the limit in which the fluid takes 
the universal form for many black holes, but our results 
show that without taking the Rindler limit, the fluid 
keeps properties of D0-branes such as the correct 
background charge. 
We have also found that when the cutoff surface is far from the horizon, the linear responses of Matrix theory do not correspond to the hydrodynamic stress tensor and current on $S^8$. Especially, we have found that in the low frequency region, the vector modes of the linear responses do not possess the pole structure, although the vector modes of the hydrodynamic stress tensor and current possess the diffusion pole. This fact does not depend on the field redefinitions of the source fields. From our results, we conclude that the linear responses of the D0-branes cannot be explained by hydrodynamics.

Three comments are in order. Firstly, we have analyzed the linear responses of Matrix theory in the AdS frame, which is not the conventional frame, such as the  Einstein frame or string frame. However, the choice of the frame does not change the dimensionless transport coefficients such as $\eta/s$, which are the physically sensible quantities.

Secondly, if we were able to absorb the extra factor $\Gamma(1+\frac{l}{7})^4/\Gamma(1+\frac{2l}{7})^2$  in (\ref{eq:finaltensoru1}) into the field redefinition of $\bbar$, the tensor mode of the linear responses for $u_c\simeq 0$ would take the same form as the hydrodynamic stress tensor. However, since the discussion of the pole structure in the vector modes is independent of the field redefinitions, our conclusion does not change.

Finally, to understand what occurs in the D0-branes in the time-dependent external sources, we also need to analyze the linear responses in the high frequency region (or full  frequency region). If we obtain the linear responses of the D0-branes in the high frequency region, we might be able to discuss the fast scrambling time via the gauge/gravity correspondence \cite{Sekino:2008he,Susskind:2011ap}. This should be investigated in a  future work.

\section*{Acknowledgments}
We would like to thank to Satoshi Iso, Kazutoshi Ohta, and Takeshi Morita  for useful  discussions and comments.  
The work of Y.M. is supported in part by JSPS Research Fellowship for Young Scientists and 
Grant-in-Aid for JSPS Fellows No.\ 23-2195. 
The work of Y.S. is supported in part by JSPS Grant-in-Aid for
Scientific Research (A) No. 23244057 and (C) No. 24540293.

\appendix
\section{Spherical harmonics on $S^8$} \label{sec:harmonics}
We briefly summarize the definitions and  properties of the spherical harmonics on $S^8$ according to \cite{Sekino:1999av}.

A scalar field $\phihat$ on $S^8$ with radius $R$ can be expanded as 
\begin{align}
\phihat=\sum_I \varphi^I(R)Y^I(x^i),
\end{align}
where $I$ denotes the angular momentum indices and $x^i~(i=1,\cdots, 8)$ are the spherical coordinates on the sphere. 
The function $Y^I$ is called the scalar harmonic.  In terms of the normalized Cartesian coordinates $\{x^m|m=1,\cdots, 9,~ x^mx_m=1 \}$, the explicit form of $Y^I$ is given by
\begin{align}
Y^I=C^I_{m_1\cdots m_l}x^{m_1}\cdots x^{m_l},~~~~~(l=0, 1, \cdots),
\end{align}
where  $C^I_{m_1\cdots m_l}$ are totally symmetric and traceless in the indices $(m_1, \cdots, m_l)$. The scalar harmonic satisfies
\begin{align}
\nabla^i\nabla_i Y^I=-\frac{l(l+7)}{R^2}Y^I,~~~~~~(l\geq 0),
\end{align}
where $\nabla_i $ is the covariant derivative on the sphere.

A vector field $\Ahat_i$ on the sphere can be expanded as
\begin{align}
\Ahat_i=\sum_I a^I(R)Y^I_i(x^i)+\sum_I \bar{a}^I(R)\nabla_i Y^I(x^i). \label{eq:vectorexp}
\end{align}
The function $Y^I_i$, which is divergentless $\nabla_iY_I^i=0$, is called the vector harmonic. In terms of the normalized Cartesian coordinates, the explicit form is given by
\begin{align}
Y^I_n=C^I_{nm_1\cdots m_l}x^{m_1}\cdots x^{m_l},~~~~(l=1,2,\cdots),
\end{align}
where the coefficients $C^I_{nm_1\cdots m_l}$ are antisymmetric under the exchange of $n$ and $m_1$ and totally symmetric and traceless with respect to the indices $(m_1, \cdots, m_l)$. The vector harmonic satisfies
\begin{align}
\nabla_i\nabla^i Y^I_j&=\frac{-l(l+7)+1}{R^2}Y^I_j,~~~~~~(l\geq 1).
\end{align}
If we impose the gauge condition $\nabla_i\Ahat^i=0$, the second term of (\ref{eq:vectorexp}) vanishes. Then, the  harmonic expansion is simplified as follows:
\begin{align}
\Ahat_i=\sum_I a^I Y^I_i.
\end{align}

A symmetric traceless tensor on the sphere can be expanded as 
\begin{align}
h_{ij}-\frac{1}{8}g_{ij}h^k_k&=\sum_I b^I(R)Y^I_{ij}(x^i)+\sum_I \bar{b}^I(R)(\nabla_i Y^I_j+\nabla_j Y^I_i)(x^i) \notag \\
&+\sum_I \bar{\bar{b}}^I(R)(\nabla_i\nabla_j-\frac{1}{8}g_{ij}\nabla^k\nabla_k)Y^I(x^i).
\end{align}
The function $Y^I_{ij}$, which is symmetric, traceless and divergentless, is called tensor harmonic. In terms of the normalized Cartesian coordinates, the explicit form is given by
\begin{align}
Y^I_{n_1n_2}=C^I_{n_1n_2m_1\cdots m_l}x^{m_1}\cdots x^{m_l},~~~~(l=2,3,\cdots),
\end{align}
where the coefficients $C^I_{n_1n_2m_1\cdots m_l}$ are antisymmetric under the exchange of $(n_1,m_1)$, symmetric under the exchange of $(n_1,n_2)$,  and  totally symmetric and traceless with respect to $m_1, \cdots, m_l$. The tensor harmonic satisfies
\begin{align}
\nabla_i\nabla^i Y^I_{jk}&=\frac{-l(l+7)+2}{R^2}Y^I_{jk},~~~~~~(l\geq 2).
\end{align}
If we impose the gauge condition $\nabla^i(h_{ij}-\frac{1}{8}g_{ij}h^k_k)=0$, the harmonics expansion is simplified as follows:
\begin{align}
h_{ij}-\frac{1}{8}g_{ij}h^k_k=\sum_I b^IY^I_{ij}.
\end{align}

\section{Derivation of on-shell action} \label{sec:on-shell}
\subsection{Tensor mode} \label{sec:tensoronshell}
We insert 
\begin{align}
g_{\mu\nu}&=\gbar_{\mu\nu}+h_{\mu\nu}, \\
A_{\mu}&=\Abar_{\mu}+\Ahat_{\mu}, \\
\phi&=\phibar+\phihat,
\end{align}
into the action (\ref{eq:IIA'}) and expand the action around the background fields up to the quadratic order of the perturbations. Using the mode expansions (\ref{eq:scalarmode})--(\ref{eq:tensormode}) and the formulas for the spherical harmonics in  Appendix \ref{sec:harmonics}, we obtain for the tensor mode,
\begin{align}
2\kappa^2S'_{IIA}&=2D_2\int d^2x \sqrt{-\gbar_2}\Phi \bigg[ \bigg(-\frac{63}{25}-\frac{1}{25}l(l+7)\bigg)\Rt^{-2}b^2 +b\nabla_a\nabla^ab +\frac{3}{4}\nabla_ab \nabla^a b\bigg], \label{eq:baction}
\end{align}
where $a,b=0,z$; $\Phi=e^{-\frac{6}{7}\phibar}=(\Rt/z)^{\frac{9}{5}}$; $\sqrt{-\gbar_2}$ is the square root of the determinant of $\gbar_{ab}$; and $n^a$ is the unit normal  to the cutoff surface. By varying the action (\ref{eq:baction}) with respect to $b$, we find the covariant form of the equation of motion (\ref{eq:eomb}),
\begin{align}
0&=\Phi \nabla_a\nabla^a b+\nabla_a\Phi \nabla^a b+2\nabla^a\nabla_a\Phi b-\frac{4}{25}\Rt^{-2}\Phi(63+l(l+7))b. \label{eq:covbeq}
\end{align}
Inserting the equation of motion (\ref{eq:covbeq}) into the action (\ref{eq:baction}), we find
\begin{align}
2\kappa^2S'_{IIA}&=2D_2\int_{z=z_c} dt \sqrt{-\gbar_{00}}\bigg[-\frac{3}{4}n^a\Phi b\nabla_ab+\frac{1}{2}n^a \nabla_a\Phi b^2\bigg] \notag \\
&=-\frac{3}{2}\Rt^{\frac{9}{5}}D_2\int_{z=z_c}dt z^{-\frac{9}{5}} f b\del_z b-\frac{9}{5}\Rt^{\frac{9}{5}}D_2\int_{z=z_c}dt z^{-\frac{14}{5}}fb^2.
\end{align}

Besides, we have to add the Gibbons-Hawking term (\ref{eq:GB}) to the action.  For the tensor mode, the Gibbons-Hawking term is
\begin{align}
2\kappa^2S_{GB}&=-\Rt^{\frac{9}{5}}D_2\int_{z=z_c}dtz^{-\frac{14}{5}}fb^2 +\frac{1}{2}\Rt^{\frac{9}{5}}D_2\int_{z=z_c}dt z^{-\frac{9}{5}}f'b^2+2\Rt^{\frac{9}{5}}D_2\int_{z=z_c}dt z^{-\frac{9}{5}}fb\del_zb.
\end{align}
Therefore, the total on-shell action  is
\begin{align}
2\kappa^2S_{\mathrm{on-shell}}&=2\kappa^2(S_{IIA}'+S_{GB}) \notag \\
&=\frac{1}{2}\Rt^{\frac{9}{5}}D_2\int_{z=z_c}dt z^{-\frac{9}{5}}fb\del_zb-\frac{7}{5}\Rt^{\frac{9}{5}}D_2\int_{z=z_c}dtz^{-\frac{14}{5}}(1+f)b^2, 
\end{align}
which is the same as (\ref{eq:onshellactionb}).
\subsection{Vector modes} \label{sec:vectoronshell}
In the same way, we calculate the on-shell action for the vector modes. Up to the quadratic order of the vector modes, the action is
\begin{align}
2\kappa^2S'_{IIA}&=D_1\int d^2x\sqrt{-\gbar_2} e^{-\frac{6}{7}\phi}\bigg[\bigg(-\frac{18}{25}f-\frac{47}{25}-\frac{2}{25}l(l+7)\bigg)\Rt^{-2}b^ab_a \notag \\
&-b^a\nabla_a\nabla_bb^b-(\nabla_a b^a)^2 -b^a\nabla_b\nabla_ab^b +2b^a\nabla_b\nabla^bb_a \notag \\
&-\frac{1}{2}\nabla_bb_a\nabla^ab^b +\frac{3}{2}\nabla_bb_a\nabla^bb^a+\frac{16}{49}(\del_a\phibar \del_b\phibar) b^ab^b -\frac{14}{5}g_s\Rt^{-\frac{14}{5}}z^{\frac{9}{5}}\epsilon^{ab}b_a\nabla_ba \notag \\
&-\frac{g_s^2}{2}\Rt^{-\frac{18}{5}}z^{\frac{18}{5}}\nabla_a a\nabla^a a-\frac{2}{25}g_s^2z^{\frac{18}{5}}\Rt^{-\frac{28}{5}}(l+1)(l+6)a^2\bigg], \label{eq:vectact}
\end{align}
where $\epsilon^{0z}=-(-\gbar_2)^{-1/2}=-\Rt^{-2}z^2,~ \epsilon_{0z}=(-\gbar_2)^{1/2}=\Rt^2 z^{-2}$, and $\nabla_c  \epsilon^{ab}=0$.
By varying the action with respect to $b^a$ and $a$, we obtain the covariant forms of the equations of motion (\ref{eq:eqofb0})--(\ref{eq:eqofa}),
\begin{align}
0&=\bigg(-\frac{94}{25}-\frac{36}{25}f-\frac{4}{25}l(l+7)\bigg)\Rt^{-2}\Phi b^a +\frac{32}{49}(\del^a\phibar \del_b\phibar) \Phi b^b \notag \\
&-\nabla^a\nabla_b\Phi b^b-\nabla_b\nabla^a\Phi b^b+2\nabla_b\nabla^b \Phi b^a -\nabla_b\Phi \nabla^a b^b+\nabla_b \Phi \nabla^b b^a \notag \\
&-\Phi\nabla_b\nabla^a b^b+\Phi\nabla_b\nabla^b b^a -\frac{14}{5}g_s\Rt^{-1}\epsilon^{ab}\nabla_b a, \\
0&=\frac{14}{5}g_s^{-1}\Rt^{\frac{14}{5}}z^{-\frac{19}{5}}\epsilon^{ab}\nabla_bb_a+\Rt^{2}z^{-\frac{19}{5}}\nabla_a(z^{\frac{9}{5}}\nabla^a a)-\frac{4}{25}(l+1)(l+6) z^{-2}a.
\end{align}
Inserting these equations into the action (\ref{eq:vectact}), we find
\begin{align}
2\kappa^2S'_{IIA}&=D_1\int_{z=z_c} dt \sqrt{-g_{00}}~n_a\bigg[\Phi b^a\nabla_b b^b+\frac{1}{2}\Phi b^b\nabla_b b^a-\frac{3}{2}\Phi b_b\nabla^a b^b \notag \\
&-\frac{7}{5}g_s\Rt^{-1}\epsilon^{ab}b_b a+\frac{g_s^2}{2}\Rt^{-\frac{9}{5}} z^{\frac{9}{5}} a\nabla^a a+\nabla^a\Phi  b_bb^b-\nabla_b\Phi b^bb^a\bigg] \\
&=\Rt^{\frac{19}{5}}D_1\int_{z=z_c}dt \bigg[z^{-\frac{19}{5}}b^z\del_0 b^0-z^{-\frac{24}{5}}(b^z)^2+\frac{1}{2}f^{-1}f'z^{-\frac{19}{5}}(b^z)^2+\frac{1}{2}z^{-\frac{19}{5}}b^0\del_0b^z \notag \\
&-\frac{1}{5}z^{-\frac{24}{5}}f^2 (b^0)^2+z^{-\frac{19}{5}}ff'(b^0)^2+\frac{3}{2}z^{-\frac{19}{5}}f^2 b^0\del_z b^0\bigg] \notag \\
&+\frac{7}{5}g_s\Rt D_1\int_{z=z_c} dt z^{-2}fb^0 a +\frac{g_s^2}{2}\Rt^{-\frac{9}{5}} D_1\int_{z=z_c} dt fz^{\frac{9}{5}}a\del_z a.
\end{align}
Since the Gibbons-Hawking term for the vector modes is
\begin{align}
2\kappa^2S_{GB}&=\Rt^{\frac{19}{5}}D_1\int_{z=z_c} dt \bigg[\bigg(3z^{-\frac{24}{5}}f^2-\frac{3}{2}z^{-\frac{19}{5}}ff'\bigg)(b^0)^2-2z^{-\frac{19}{5}}f^2b^0\del_zb^0\bigg],
\end{align}
the total on-shell action for the vector modes  is
\begin{align}
2\kappa^2S_{\mathrm{on-shell}}&=\Rt^{\frac{19}{5}}D_1\int_{z=z_c}dt \bigg[\frac{1}{2}z^{-\frac{19}{5}}b^z\del_0 b^0+\frac{2}{5}z^{-\frac{24}{5}}(b^z)^2-\frac{7}{5}f^{-1}z^{-\frac{24}{5}}(b^z)^2 \notag \\
&+\frac{7}{5}z^{-\frac{24}{5}}f^2 (b^0)^2+\frac{7}{5}z^{-\frac{24}{5}}f (b^0)^2-\frac{1}{2}z^{-\frac{19}{5}}f^2 b^0\del_z b^0\bigg] \notag \\
&+\frac{7}{5}g_s\Rt D_1\int_{z=z_c} dt z^{-2}fb^0 a +\frac{g_s^2}{2}\Rt^{-\frac{9}{5}} D_1\int_{z=z_c} dt fz^{\frac{9}{5}}a\del_z a.
\end{align}

\end{document}